\documentclass[12pt,preprint]{aastex}

\newcommand {\ea} {{\it et~al.}}
\newcommand {\be} {\begin{equation}}
\newcommand {\ee} {\end{equation}}

\shorttitle{ 
KLEIN-NISHINA EFFECTS IN NON-THERMAL SOURCES}

\shortauthors{WHO \ea}

\begin{document}


\title{Klein-Nishina Effects in the Spectra of Non-Thermal 
Sources Immersed in External Radiation Fields}

\author{Rafa{\l} Moderski \altaffilmark{1},
Marek Sikora \altaffilmark{1},
Paolo S. Coppi \altaffilmark{2,3}, and
Felix Aharonian \altaffilmark{3}
}

\altaffiltext{1}{Nicolaus Copernicus Astronomical Center,
Bartycka 18, 00-716 Warsaw, Poland; \tt{moderski@camk.edu.pl}}
\altaffiltext{2}{Yale University, New Haven, CT 06520-8101, USA}
\altaffiltext{3}{Max-Planck-Institut f\"ur Kernphysik, Heidelberg, Germany}

\begin{abstract}
 
We provide a systematic numerical and analytical 
study of Klein-Nishina (KN) effects
in the spectrum produced by a steady state,
non-thermal source where rapidly accelerated
electrons 
cool by emitting synchrotron radiation and Compton upscattering ambient
photons produced outside the source.
We focus on the  case where
$q,$ the ratio of the ambient radiation field 
to source magnetic field energy densities, significantly exceeds unity.
We show that the KN reduction in the electron Compton
cooling rate causes the steady-state electron spectrum
to harden at energies $\gamma \gtrsim \gamma_{KN},$ 
where
$\gamma_{KN}= 1/4\epsilon_0$ and $\epsilon_0=h\nu_0/m_ec^2$ is the
characteristic ambient photon energy.
This hardening becomes noticeable in the
synchrotron radiation from electrons with energies as low as
$0.1 \gamma_{KN} $ and changes the synchrotron
spectral index relative to its Thomson limit value by 
as much as $\Delta \alpha \sim 0.75$
for $q\gg 1.$  The source synchrotron spectrum thus shows
a high-energy ``bump'' or
excess even though the electron acceleration spectrum has no such excess.
In contrast, the low-energy Compton gamma-ray spectrum shows little
distortion because the electron hardening compensates for the
KN decline in the scattering rate. For sufficiently high
electron energies, however, Compton cooling becomes
so inefficient that synchrotron cooling dominates -- an effect
omitted in most previous studies. The hardening of the electron
distribution thus stops, leading to a rapid
decline in Compton gamma-ray emission, i.e., a strong spectral break
whose location does not depend on the maximum electron energy.
This break can limit the 
importance of Compton gamma-ray pair production on ambient photons
and implies that a source's synchrotron luminosity may
exceed its Compton luminosity even though $q > 1$.
We discuss the importance
of these KN effects in blazars, micro-quasars, and pulsar binaries.

\end{abstract}


\section{INTRODUCTION}

Probably because of the initially greater sensitivity of radio
telescopes, studies of the emission from non-thermal sources,
e.g., the powerful radio jets found in AGN, focused on synchrotron radiation
from relativistic electrons moving in a magnetic field. It was quickly
realized, however, that the same relativistic electrons would also Compton
upscatter any ambient low-energy photons to produce emission at much
higher, e.g., X-ray and gamma-ray, energies. A classic discussion of
the emission spectrum expected from a gas of relativistic electrons where
synchrotron radiation and Compton scattering are  the dominant 
energy loss mechanisms may be found, for example, 
in Felten \& Morrison (1966). Several of the standard approximations still 
used today are presented there, e.g., the delta function approximation 
which provides a one-to-one relation between the energies of the electrons and 
the synchrotron and Compton upscattered photons produced by them.
Of particular interest, those authors note that the processes of 
Compton upscattering in the Thomson regime 
and synchrotron radiation
both lead to quasi-continuous electron energy losses that are proportional to
the square of the electron Lorentz factor. (This is in fact 
not a surprise given that  synchrotron radiation
may be viewed as the Compton upscattering of virtual magnetic 
field photons [see Blumenthal \& Gould 1970].) This implies
that the shape of the energy distribution of cooled 
electrons does not depend on which process dominates their
cooling and that the ratio of the luminosities for the 
resulting Compton and
synchrotron emission components simply goes as 
$L_{C}/L_{syn} = u_0/u_B,$ where $u_0$ and $u_B$ are
respectively the (co-moving) low-energy radiation and 
magnetic field energy densities inside the source.
This is very useful to know, for example, since if one   
measures $L_{C}$ and $L_{syn}$ and knows
$u_{0},$ e.g., if the relevant low-energy ambient photons are due
to the cosmic microwave background, then one can immediately
derive the magnetic field strength in the source.  
Also, the shapes of the synchrotron and Compton emission spectra
will be similar,
with the Compton spectrum simply shifted up in energy relative
to the synchrotron one by the
factor $\epsilon_0/\epsilon_B$ where 
$\epsilon_B = B/B_{cr},$ $B_{cr}=4.4\times 10^{13}$ G, and 
$\epsilon_0 = h\nu_0/m_ec^2$ is the typical ambient photon energy.
(For mono-energetic ambient photons,
the spectral shapes are in fact identical except for the effects of 
self-absorption in the low-energy portion of the synchrotron spectrum.)

It is important to remember, however, that all of
these convenient rules of thumb, used
in many interpretation papers,  break down when the
electrons are sufficiently energetic to scatter with 
ambient photons in the Klein-Nishina (KN) limit, i.e.,  when 
$\gamma > \gamma_{KN}=1/4\epsilon_0$ (e.g., Blumenthal \& Gould 1970).
In this case, the relation between
upscattered  and incident photon energy changes (the upscattered
photon cannot have more energy than the incident electron),
electrons can lose most of their energy in a single scattering
rather than cooling quasi-continuously as a result of many small energy
losses, and to the extent that a cooling rate is relevant, the
ratio of Compton to synchrotron energy losses is now
a decreasing function of energy and is less than $u_0/u_B.$
A careful re-examination of the implications of treating
the Compton scattering process 
correctly in this limit is 
timely now that satellite and ground-based gamma-ray 
telescopes have shown many extragalactic and some
galactic compact objects to be strong
emitters of  GeV and TeV radiation. If produced by Compton
up-scattering, i.e., the Inverse Compton (IC) process, 
then these gamma-rays likely result from scatterings in 
the KN regime, and we must be accordingly careful
in the interpretation of the multi-wavelength observations for these
objects. 

Some of the possible KN ``corrections'' have
already been discussed or are well-known. 
An obvious limit is the
case when the ambient (target) photon energy density in the 
source is small and the energy density ratio 
$q=u_0/u_B$ is thus $\ll 1.$ In this case, 
synchrotron losses always dominate over Compton
losses regardless of electron energy.  If we have a source
where electrons are rapidly accelerated to high energies and
subsequently cool via radiative losses, then the steady state
spectrum of the source (assuming on-going acceleration) is
determined solely by the magnetic field and the ``injection'' energy spectrum
of electrons produced by the rapid acceleration process.
We can therefore consider the electron distribution to be fixed,
e.g., a power law with $n_\gamma \propto
\gamma^{-s},$ 
and just need to carry out the relatively easy
computation of the Compton spectrum upscattered by such an 
electron power law. For simplicity, let us assume that the ambient radiation
field is approximately mono-energetic, again with 
characteristic energy $\epsilon_0.$
In this case, the resulting Compton emissivity is approximately
a  broken power law,
with the usual Thomson-limit result of $j_\epsilon \propto 
\epsilon^{-(s-1)/2}$ for $\epsilon \ll \epsilon_{KN},$ and the extreme KN-limit
result of $j_\epsilon \propto \epsilon^{-s} \log(\epsilon)$ for 
$\epsilon \gg \epsilon_{KN}$ where $\epsilon_{KN} = \gamma_{KN},$
e.g., see Aharonian \& Atoyan (1981). In other words, the synchrotron
spectrum is unchanged from a Thomson-approximation calculation
while the Compton spectrum shows a strong spectral break at $\epsilon
\sim \epsilon_{KN}$ with a change in spectral index 
$\Delta \alpha = (s+1)/2.$ For $s \gtrsim 2,$ which seems typical
for the known very high-energy gamma-ray sources, $\Delta \alpha \gtrsim 1.5,$
i.e., is large, and given the usually poor statistics at the highest
energies, this spectral break can easily be misinterpreted as a 
an exponential cutoff due to the maximum electron
energy in the source. 

A much less obvious but still important limit is the
case $q \gg 1.$  For $u_0$ sufficiently large,
one can in principle go the opposite limit 
and ignore the effects of synchrotron cooling,
solving directly the kinetic equation for the evolution
of electrons due to Compton scattering.
The steady state
electron and upscattered photon spectra obtained
for continuous electron acceleration and complete electron cooling
via Compton scattering 
are discussed in detail by Zdziarski (1989) 
(see also Zdziarski \& Krolik, 1993).
The main conclusion is that
the equilibrium electron distribution hardens for $\gamma>\gamma_{KN}$
compared to the Thomson approximation result. This is because
electron energy losses are relatively less efficient in the KN
limit and electrons remain longer at higher energies.  This hardening
of the electron spectrum compensates the decreased efficiency
of Compton upscattering, and the resulting
photon spectrum shows {\it no} ``Klein-Nishina
break'' at $\epsilon\sim {\epsilon}_{IC,KN}.$.
Interestingly, for a rapid acceleration
process that effectively injects electrons into the source with 
an energy distribution $Q(\gamma) \propto \gamma^{-p}$ with
$p=2,$ the Compton emissivity is $j_\epsilon \propto \epsilon^{-2},$
i.e., exactly the answer that would have been obtained in the Thomson limit.
Note, though, an important difference from the Thomson result is 
that the correctly computed spectrum always cuts off at $\epsilon
\sim \gamma_{max},$  the maximum electron Lorentz factor, 
and that this  cutoff
is independent of the target photon energy $\epsilon_0.$
Also, for $p\ne 2$, the Compton spectrum in the KN regime does not follow
the slope of the spectrum in the Thomson regime, 
steepening a bit if $p>2$, and hardening if $p<2$. 

The analysis just mentioned, however, do not include synchrotron 
losses and do not consider the synchrotron emission produced
as a result of these losses. (The condition $q \gg 1,$ in fact only 
guarantees
that synchrotron losses are negligible for $\gamma <\gamma_{KN}.$
At higher energies, the Compton loss rate decreases due to 
KN effects and eventually synchrotron cooling always dominates.)
They thus miss several effects that are potentially important
for high energy sources. 
First, the hardening of electron distribution in the KN regime 
leads to a hardening of co-spatially produced synchrotron radiation.
Prior work which touches on this aspect includes:
 Dermer \& Atoyan (2002) which invokes KN
effects to help explain 
the production of X-rays in large scale jets via synchrotron
radiation;  Ravasio et al. (2003) which invoke KN
effects to explain the spectral ``glitch;'' Kusunose, Takahara \&
Kato (2004) and Kusunose \& Takahara (2005) which study KN effects 
in the context of FSRQ (flat-spectrum-radio-quasars).
Secondly, for sufficiently high electron energies, synchrotron losses dominate
and the electron distribution ``saturates'' to one with the same 
slope as in the Thomson regime, but with a normalization factor $q$ times 
larger. This effect was  included 
by Khangulian \& Aharonian (2005) in their studies 
of outflows from compact objects in high mass 
X-ray binary systems, but not by Kusunose \& Takahara (2005),
even though some of their calculations involve electron energies
high enough for synchrotron losses to be important.

In this paper, we provide a systematic study of KN effects in 
steady-state sources with $q\gg 1$, covering all the effects mentioned
above. We use accurate numerical techniques that solve the
exact integro-differential equations for the steady-state photon
and electron energy distributions. However, we also  
present approximations that allow one to follow the various effects 
analytically using simple algebraic functions.  We focus our studies on
the case where the ambient radiation field is dominated by external 
photon sources
with mono-energetic or power-law spectra. 
The paper is organized as follows. In \S 2 we analyze inverse-Compton 
electron energy losses in the KN regime and compare them with 
the corresponding 
synchrotron energy losses. In \S 3 we present general and approximate
formulas for the inverse-Compton emissivities for the cases of 
isotropic and  beamed ambient radiation fields. The energy 
distribution of relativistic electrons in a steady-state source
with continuous electron injection
is discussed in \S 4. The KN effects in the 
synchrotron and inverse-Compton spectra produced by such a source 
are analyzed in \S 5. Our results are discussed in the context of 
specific astrophysical sources in \S 6 and summarized in \S 7.

\section{ELECTRON ENERGY LOSSES}

\subsection{Inverse Compton energy losses}

The rate of inverse Compton (IC) energy losses of relativistic, isotropically 
distributed  electrons is (see Appendix A)
\be \vert \dot \gamma \vert_{IC}  =
 {4c \sigma_T \over 3 m_e c^2} u_0 \gamma^2 F_{KN} \, , \label{dotg3}\ee
where
\be F_{KN}= {1 \over u_0} 
\int_{\epsilon_{0,min}}^{\epsilon_{0,max}}
 f_{KN}(\tilde b) u_{\epsilon_0} d \epsilon_0  \, ,\label{FKN} \ee
$u_0=\int_{\epsilon_{0,min}}^{\epsilon_{0,max}}
 u_{\epsilon_0} d\epsilon_0$ is the total energy
density of the radiation field, $u_{\epsilon_0}$ is the energy  
distribution of the ambient photons, and we re-express the
electron energy as  
$\tilde b=4 \gamma \epsilon_0$ noting 
$\tilde b=1$ corresponds to 
the transition between the Thomson and KN scattering
regimes.
The function
$f_{KN}(\tilde b)$ is given by Eqs.~(\ref{fKN}) and (\ref{gb}) in Appendix A.
For $\tilde b \ll 1$ (Thomson limit), $f_{KN} \simeq 1$; for $\tilde b\gg1$ 
(KN limit), $f_{KN} \simeq (9/(2\tilde b^2))(\ln\tilde b - 11/6)$.
For
$ b \lesssim 10^4,$   
$f_{KN}(\tilde b)$ can be approximated by 
\be f_{KN} \simeq {1 \over (1+\tilde b)^{1.5}} \, . \label{fKNap} \ee

To estimate the effects
of having an extended target photon energy distribution,  
let us assume that $u_{\epsilon_0}$ is a power law $
\propto \epsilon_0^{-\alpha}$ and that
$b_{max} = 4\epsilon_{0,max} \gamma_{max} < 10^4,$ so we may use
approximation (\ref{fKNap}). Then we may write,
\be 
F_{KN} \propto \int_{\epsilon_{0,min}}^{\epsilon_{0,KN}}
\epsilon^{-\alpha_0} d\epsilon_0 +
\epsilon_{0,KN}^{1.5}\int_{\epsilon_{0,KN}}^{\epsilon_{0,max}}
\epsilon_0^{-\alpha_0-1.5}d\epsilon_0, \label{FKNap1} 
\ee  
where $\epsilon_{0,KN} = 1/4\gamma.$ KN effects become important
and $F_{KN} < 1$ for $\gamma> \gamma_{KN} \equiv 1/(4\epsilon_{0,max}).$
For $\alpha_0 < -0.5,$  Compton energy losses for 
electrons with $\gamma>\gamma_{KN}$ are dominated by scatterings
on photons with the highest energies, $\sim \epsilon_{0,max}.$
We may then treat the photon distribution as mono-energetic, with 
$F_{KN} \simeq f_{KN}(\tilde b=4\gamma\epsilon_{0,max}).$
For $-0.5< \alpha_0 <1,$ Compton losses are instead dominated by
scattering on photons with energy $\sim \epsilon_{0,KN}.$ In this
case, we may make the so-called ``Thomson-edge'' or
``Klein-Nishina cutoff'' approximation that 
\be
F_{KN} \simeq {\int_{\epsilon_{0,min}}^{\epsilon_{0,KN}}
\epsilon_0^{-\alpha_0} d\epsilon_0 \over 
\int_{\epsilon_{0,min}}^{\epsilon_{0,max}}
\epsilon_0^{-\alpha_0} d\epsilon_0} \simeq
(\epsilon_{0,KN}/\epsilon_{0,max})^{-\alpha_0+1} = b^{\alpha_0-1},
\label{FKNedge_0}
\ee
if rewrite the 
electron energy as $b=4\epsilon_{0,max}\gamma$ and 
have $\epsilon_{0,min}\ll\epsilon_{0,max}.$ A useful
approximation that interpolates to the Thomson limit for $b\to 0$
and has good accuracy for $0<\alpha_0<1$ is then given by,
\be F_{KN} \simeq {1 \over (1+b)^{1-\alpha_0}}. \label{FKNedge} \ee
If we instead have $\alpha_0 > 1,$ the Compton losses of the electrons
are dominated by scatterings on the lowest energy photons available, 
at $\sim \epsilon_{0,min}.$ We may then use the approximation,
$F_{KN} \simeq
f_{KN}(\tilde b=4\gamma\epsilon_{0,min}).$ 
If we further have $4\gamma_{max}\epsilon_{0,min} < 1,$
then in fact all scatterings are effectively in the Thomson regime
even if higher energy ambient photons are present and 
$b_{max} > 1.$
To gauge the accuracy of approximations (\ref{fKNap}) and
(\ref{FKNedge}), we compare in  
Fig.~\ref{fig1} the approximate values of $F_{KN}(b)$ 
to the exact ones for three different ambient
photon spectra: mono-energetic, and power laws with     
$\alpha_0 \equiv -d\ln u_{\epsilon_0}/d\ln \epsilon_0 =0.0$ 
and $0.5$. 
From an analysis similar to that for
the power law case, one can show that 
$F_{KN}$ for a Planckian (thermal) 
distribution is well-approximated by treating
the Planckian as a  mono-energetic
photon distribution with energy $2.8kT.$

\bigskip

\subsection{Synchrotron energy losses vs. inverse-Compton energy losses}

Noting that the rate of synchrotron losses in a tangled magnetic field
of strength $B$ is
\be \vert \dot \gamma \vert_{syn}  =
 {4c \sigma_T \over 3 m_e c^2} u_B \gamma^2 \ee
where $u_B=B^2/8\pi$ is the 
magnetic energy density, we have (see Eq.\ref{dotg3}) 
\be {\dot \gamma_{IC} \over \dot \gamma_{syn}}
= q F_{KN} \, \ee
where $q \equiv u_0/u_B$. Since $F_{KN} < 1$ for any value of $b$, 
for $q <1$ 
the energy losses for all electrons are dominated by synchrotron radiation.
For $q >1$, the energy losses for electrons with $\gamma < \gamma_s$ are 
dominated by inverse Compton scattering while losses for electrons with 
$\gamma > \gamma_s$ are instead dominated by synchrotron radiation, where 
$\gamma_s= b_s/4 \epsilon_{0,max}$, and 
$b_s$ is the solution of the equation $qF_{KN}=1$, plotted in Fig.~\ref{fig2}.
For a mono-energetic ambient photon spectrum (see Eq.\ref{fKNap})
\be b_s \simeq q^{2/3} - 1 \, , \label{bs0} \ee
while for a power-law spectrum  with $0<\alpha_0<1$ (see Eq.\ref{FKNedge}),
\be b_s \simeq q^{1/(1-\alpha_0)} -1 \, . \label{bs} \ee

The relative role of the IC and synchrotron energy losses is 
illustrated in Fig.~\ref{fig3}. We show there
\be {\dot \gamma_{IC} \over \dot \gamma_{tot}} =
{q F_{KN}(b) \over 1 + q F_{KN}(b)} \, , \label{IC/tot} \ee 
and 
\be {\dot \gamma_{syn} \over \dot \gamma_{tot}} =
{1 \over 1 + q F_{KN}(b)} \, , \label{syn/tot} \ee 
for three different spectra of the ambient radiation field,
assuming that $\dot \gamma_{tot}= \dot \gamma_{IC}+\dot \gamma_{syn}$.

\section{INVERSE-COMPTON EMISSIVITY}

The inverse Compton emissivity for isotropically distributed electrons
is 
\be j_{\epsilon(IC)} = 
\int {{\partial P_{\epsilon(IC)}(\gamma) \over 
\partial \Omega} n_{\gamma} d\gamma}  \, , \label{jIC} \ee
where 
\be {\partial P_{\epsilon(IC)}(\gamma) \over \partial \Omega}=
{\partial \dot N_{sc}(\epsilon,\gamma) \over \partial \epsilon \partial \Omega}
\, \epsilon m_ec^2 \, , \label{dPeps} \ee
is the IC-power per unit photon energy per solid angle per electron.
For two particular cases, mono-directional beams of target photons and
an isotropic radiation field, the respective scattering rates are
given by Eqs.~(\ref{Nsctheta}) and (\ref{Nsc}). Inserting them into 
Eq.~(\ref{jIC}) gives,
for the beamed mono-energetic external radiation field,  
\be \epsilon j_{\epsilon(IC)}(\theta) =
{3\over 16\pi} c \sigma_T u_0 \left(\epsilon \over \epsilon_0 \right)^2 
\int {n_{\gamma} \over {\gamma}^2}  
f(\gamma, \epsilon, \epsilon_0, \theta)  d\gamma  
\, , \label{jICt} \ee
and for the isotropic ambient radiation field,
\be \epsilon j_{\epsilon(IC)} =
{3\over 16\pi} c \sigma_T  u_0 \left(\epsilon \over \epsilon_0 \right)^2 
\int {n_{\gamma} \over {\gamma}^2} 
f_{iso}(\gamma, \epsilon, \epsilon_0) d\gamma 
\, , \label{jICi} \ee
where $\theta$ is the scattering angle 
(the angle between the photon beam
and the direction to the observer), the functions
$f(\gamma, \epsilon, \epsilon_0, \theta)$ and
$f_{iso}(\gamma, \epsilon, \epsilon_0)$ are given by Eqs.~(\ref{ftheta}) and
(\ref{fiso}).
These formulas can easily be generalized for any spectrum of external 
radiation field by replacing $u_0$ by $u_{\epsilon_0} d\epsilon_0$
and integrating them over $\epsilon_0$.

A useful first approximation for the IC spectrum (e.g., see Coppi \& 
Blandford 1990) may be obtained by making the delta function 
approximation $ f(\epsilon,\gamma,\epsilon_0) \propto 
\delta[\epsilon - \bar\epsilon_{(IC)}(\gamma)],$
where
\be \bar\epsilon_{IC}(\gamma,\theta) = 
{\int \epsilon f(\gamma, \epsilon, \epsilon_0, \theta) 
d\epsilon
\over \int f(\gamma, \epsilon, \epsilon_0, \theta) d\epsilon} \, 
\label{barepst} \ee
and 
\be \bar\epsilon_{IC}(\gamma)
 = {\int \epsilon f_{iso}(\gamma, \epsilon, \epsilon_0) 
d\epsilon
\over \int f_{iso}(\gamma, \epsilon, \epsilon_0) d\epsilon} \, \label{bareps}
\ee
are the  average energies of photons produced by the scattering of 
photons of energy $\epsilon_0$ by electrons with energy $\gamma.$
A useful quantity to know when dealing with the KN regime
is the electron's scattering inelasticity, i.e., 
the average fraction of the incident electron's energy that is transferred 
to the scattered photon: 
${\cal A}=\bar \epsilon_{IC}(\gamma)/\gamma.$ This is shown in Fig.~\ref{fig4}
for the case of a mono-energetic ambient photon field. 

Inserting the delta function approximation into
 Eqs. (\ref{jICt}), and (\ref{jICi}),
we can rewrite the  IC emissivities as
\be \epsilon j_{\epsilon(IC)}(\theta) \simeq { n_{\gamma}\gamma\over 4\pi}  
\, |\dot\gamma|_{IC} \, m_e c^2 \,\,  \chi(\gamma, \theta) \,
{d\ln \gamma \over d\ln \bar\epsilon_{(IC)}(\theta)}
 \, , \label{jICtap} \ee
and
\be \epsilon j_{\epsilon(IC)} \simeq { n_{\gamma}\gamma\over 4\pi}  
\, |\dot\gamma|_{IC} \, m_ec^2 \,
{d\ln \gamma \over d\ln \bar\epsilon_{(IC)}}
 \, , \label{jICiap} \ee
respectively,
where
\be \chi(\gamma, \theta) = {
\int f(\epsilon,\epsilon_0, \gamma, \theta) \epsilon \, d\epsilon
\over  
\int f_{iso}(\epsilon,\epsilon_0, \gamma) \epsilon \, d\epsilon}
\equiv {\bar \epsilon_{IC}(\gamma,\theta) \over 
\bar \epsilon_{IC}(\gamma)}\, .\ee 
Note that in the Thomson limit, $\bar \epsilon_{IC} \propto \gamma^2,$
so that $d \ln \gamma/d \ln \bar \epsilon_{IC} = 1/2,$ while
in the KN limit, $\bar \epsilon_{IC} \sim \gamma$ so that
$d \ln \gamma/d \ln \bar \epsilon_{(IC)} = 1.$ 

We can make an analogous approximation for the synchrotron emissivity,
provided  that the magnetic fields are isotropic and 
$\gamma_{max} < B_{cr}/(4B)$. We have then,
\be \epsilon j_{\epsilon(syn)} \simeq {n_{\gamma}\gamma \over 4\pi} \, 
|\dot\gamma|_{syn}m_ec^2 \,  {d\ln \gamma \over d\ln \bar\epsilon_{(syn)}} 
={1 \over 2} \,  {n_{\gamma}\gamma \over 4\pi} \, 
|\dot\gamma|_{syn}m_ec^2 \, \label{jsyn} 
\ee
where $\bar\epsilon_{syn}= (4/3) \gamma^2 (B/B_{cr})$ and 
$B_{cr} \equiv 2 \pi m_e^2 c^3/(he) \simeq 4.4 \times 10^{13}$Gauss. 

\section{STEADY-STATE ELECTRON ENERGY DISTRIBUTION}

Under the assumptions
 that the time scales for the acceleration of individual electrons  
are much shorter than time scales of their  energy losses, that electrons 
do not escape from the cooling region, and that they also do not suffer
adiabatic losses, the evolution of the electron energy 
distribution, $n_{\gamma}$, can be described  by the 
integro-differential equation 
(Blumenthal \& Gould 1970)
\be {\partial n_{\gamma} \over \partial t} = - {\partial \over \partial \gamma}
\left(n_{\gamma} |\dot \gamma| \right) -
n_{\gamma} \int_1^{\gamma} C(\gamma,\gamma') \, d\gamma' +
\int_{\gamma}^{\gamma_{max}} n_{\gamma} C(\gamma',\gamma) \, d\gamma' +
Q \, , \, \label{dn/dt} \ee  
where $|\dot \gamma|$ is the energy loss rate due to
loss processes that may be approximated as being continuous,
e.g., synchrotron radiation,  
$C(\gamma,\gamma')$ is the probability per unit time for 
Compton scattering of an
electron with energy $\gamma$ to energy $\gamma'$, and $Q$ is the electron
injection function, i.e. the electron production rate per unit time, per
energy and per volume. The transition rates $C(\gamma, \gamma')$ have been
derived by Jones (1968) for a mono-energetic ambient radiation field,
and by Zdziarski (1988) for power-law ambient radiation fields.

To illustrate in a simple way  
the effects of KN corrections, we will use a  continuity version of 
the kinetic equation
\be {\partial n_{\gamma} \over \partial t} = - {\partial \over \partial \gamma}
\left(n_{\gamma} |\dot \gamma|_{tot} \right) +
Q \, , \, \label{Cdn/dt} \ee  
despite  the fact that in the KN regime the fractional
electron energy losses per scattering are {\it not} negligible 
(see Fig.~\ref{fig4}).
Such a simplification can be justified by noting  
that the results obtained by 
Zdziarski (1989) show that the electron distributions obtained by
using the 
exact 
integro-differential equation versus the continuity equation  are 
qualitatively very similar
and that significant differences (up to a factor few) occur only if 
both the electron injection function and the ambient radiation field are 
mono-energetic. In the cases studied here, the differences
are further reduced by the increasingly important role of the (continuous) 
synchrotron energy losses at the highest energies, particularly if 
$b_{max} > b_s$.  We demonstrate this explicitly in Fig. 5-7 where
we compare the results of exact calculations made using the
code of Coppi (1992) with the approximate results obtained by treating 
Compton losses continuously in that code. For many applications
that do not involve fitting data and thus do not require 
very high accuracy, the
continuous energy loss approximation should suffice.

A steady-state solution of the continuity equation, 
$\partial n_{\gamma}/ \partial t=0$, is
\be n_{\gamma} = {1\over |\dot \gamma_{tot}|} \int_{\gamma} Q \, d\gamma 
={3 m_e c \over 4 \sigma_T u_B} \,
{\int_{\gamma} Q \, d\gamma \over \gamma^2(1+qF_{KN})} 
\, . \label{ng} \ee
where we assume that electron energy losses are dominated by inverse Compton
process and by synchrotron radiation, i.e. that
\be |\dot \gamma_{tot}|= |\dot \gamma_{syn}|+|\dot \gamma_{IC}|
= {4 c\sigma_T u_B \over 3 m_e c^2} \gamma^2(1 +qF_{KN}) \, . 
\label{dotgtot} \ee
%
For $b_{max} \gg b_s$ and $q \gg 1$, the steady-state then  distribution has 
two asymptotes: one
for $b \ll 1$, where  electron energy losses are dominated by Thomson
scatterings, i.e. $F_{KN}=1$; and one for $b_s \ll b \ll b_{max}$, where 
electron energy losses are dominated by synchrotron radiation, i.e.
$qF_{KN} \ll 1$. The respective equilibrium electron distributions are:
$\gamma^2 n_{\gamma} \propto \int_{\gamma} Qd\gamma/(1+q)$ and
$\gamma^2 n_{\gamma} \propto \int_{\gamma} Qd\gamma$. 
For $1 < b < b_s$, despite the decrease of the KN cross 
section, electron energy losses are still  dominated by
the IC process and  the electron distribution is 
$\gamma^2 n_{\gamma} \propto \int_{\gamma} Q d\gamma/(qF_{KN})$.

For a rapid acceleration mechanism that produces a power law
electron injection spectrum, 
$Q \propto \gamma^{-p}$, with $p>1$, the two asymptotes of 
the steady state distribution (for $b \ll 1,$ and $b \gg b_s$)
are power laws with the same spectral index,  
$s=p+1$, where $s \equiv -d\ln n_{\gamma}/d\ln \gamma$.   
For  $1 \ll b \ll b_s$, i.e.,  for
$\gamma_{KN} < \gamma < \gamma_s$,
the electron  distribution  is {\it harder} than in the asymptotic
regions, with a spectral index reaching $s=p+1 + \Delta s,$ where
$\Delta s \simeq  d\ln F_{KN} / d\ln \gamma<0$. Using the analytical
approximations for $F_{KN}$ given in \S2 and going to the limit
$q\gg 1,$ one finds that  for
ambient radiation fields with mono-energetic or very sharply peaked
photon energy distributions, $\Delta s \simeq -1.5$, while for softer external 
fields, with $\alpha_0 > 0.0$, $\Delta s \simeq \alpha_0 -1$. When
$q \lesssim 10^3,$ the maximum hardening, $\Delta s,$ of the 
electron distribution is a function of $q$ and decreases with 
decreasing $q.$ (See Fig. 8 for an example of how the corresponding
hardening of the electron synchrotron depends on $q.$)
Note
that to the extent the continuous Compton cooling approximation is a good one, 
this result is independent of the injection index $p.$ 
Of course, for $b_{max} < b_s$ the  electron
energy distribution will not have the maximum deviation possible 
from the Thomson-limit ($b\ll 1$) asymptote and $\Delta s$ may 
not reach its maximum value. In this case,  
the KN induced bump or ``excess'' at the
high energy end of the electron distribution
is correspondingly 
less prominent. This is demonstrated  
in Fig.~\ref{fig5}, where
the electron distribution is computed for four different values of $b_{max}$. 

\section{SPECTRA}

\subsection{Klein-Nishina effects} 

Examples of the  
steady-state electromagnetic spectra produced 
for the case of  power-law electron injection 
function are shown in Figs.~\ref{fig6} and \ref{fig7}.
The IC spectra are
computed assuming an isotropic 
ambient radiation field.
The characteristic photon energies marked on the figures have 
the following definitions (see Eq. {\ref{bareps}): 
$\epsilon_{IC,KN} \equiv \bar \epsilon_{IC}(\gamma_{KN});$
$\epsilon_{IC,s} \equiv \bar \epsilon_{IC}(\gamma_s);$
$\epsilon_{syn,KN} \equiv \bar \epsilon_{syn}(\gamma_{KN});$
$\epsilon_{syn,s} \equiv \bar \epsilon_{syn}(\gamma_s);$
and $\epsilon_+\equiv 2/\epsilon_{0,max},$ which
is the approximate threshold
energy for photon-photon pair production on the ambient photon distribution.

As one can see, the high energy portions of the inverse-Compton (IC) 
and synchrotron spectra behave very differently. The IC spectra
do not change very much after crossing $\epsilon_{IC,KN}.$
This is because for $q\gg 1,$ Compton scattering  is still dominating
the cooling of the electrons responsible for IC photons
at $\epsilon \gtrsim \epsilon_{IC,KN}.$ Even though those
electrons scatter in the KN
regime, the decreased efficiency of Compton scattering is compensated
by the corresponding increase in their equilibrium density,
as discussed in Zdziarski \& Krolik (1993).  As one moves
to higher photon energies, however, synchrotron cooling becomes
relatively more important for the electrons responsible these
IC photons. Eventually, when the energy of the relevant electrons
exceeds $\gamma_s,$ synchrotron cooling rapidly 
dominates and the fraction of the 
electron's energy going into the Compton component plummets. 
The result is a sharp steepening or break of the IC spectrum at energy
$\epsilon_{IC,s},$ which is {\it independent} of the maximum
electron energy $\gamma_{max} > \gamma_{s}.$

When one instead considers synchrotron emission, the hardening
of the electron distribution at $\gamma>\gamma_{KN}$ is 
directly reflected
in the synchrotron spectrum.
The result is 
a synchrotron spectrum that can harden
dramatically at $\epsilon>\epsilon_{syn,KN},$ forming
a synchrotron ``bump'' if $b_{max} < b_s.$  (Note that if $q \gg 1,$
the hardening of the synchrotron component can already be noticeable
at even lower energies, $\sim \bar \epsilon_{syn}(0.1 \gamma_{KN})$.)
For $b_{max} \gg b_s$,  the 
synchrotron spectrum at $\epsilon > \epsilon_{syn,s}$
asymptotes to a spectrum with the
same slope as the low energy (Thomson regime) asymptote but
with a normalization that is a factor $q$ higher. 
 
An interesting consequence of the very different behaviors of the high energy 
portions of  IC and synchrotron spectra is 
that for the case of  a flat electron injection spectrum 
($p<2$) with $b_{max}>b_s,$ 
the luminosity of the synchrotron peak -- located around
$\epsilon_{syn,max}\equiv\bar \epsilon_{syn}(\gamma_{max})$ -- 
will exceed the luminosity of the IC peak -- located
around  $\epsilon_{IC,s}$, no matter how large we make $q$.
This dominance of the synchrotron component, even though $q=30\gg 1,$ 
is demonstrated in Fig.~\ref{fig7}.

All the spectral features just described can be reproduced  analytically,
by using the equations
\be \epsilon j_{\epsilon(IC)} \simeq 
{m_e c^2 \over 4 \pi} \, {qF_{KN} \over 1 + qF_{KN}} \,
\gamma \int_{\gamma} Qd\gamma \,\, { d\ln \gamma \over d\ln \bar\epsilon_{IC}}
\, , \label{jIC2} \ee
and 
\be \epsilon j_{\epsilon(syn)} \simeq 
{1 \over 2} \, {m_e c^2 \over 4 \pi} \, {1 \over 1 + qF_{KN}} \,
\gamma \int_{\gamma} Qd\gamma \, ,
 \label{jsyn2} \ee
which are obtained after insertion of Eq. (\ref{ng}) into Eqs. (\ref{jICiap})
and (\ref{jsyn}), respectively.
For $\gamma_{max} \gg \gamma_s$ and $q \gg 1$, the latter implying
$b_s \gg 1$ i.e. $\gamma_s \gg \gamma_{KN}$, the IC and synchrotron spectra
can be characterized as the superposition of three components
produced by electrons with: 
$\gamma \ll \gamma_{KN}$; $\gamma_{KN} \ll \gamma \ll \gamma_s$; and
$\gamma_s \ll \gamma \ll \gamma_{max}$. These components are:   
\be
{4\pi \epsilon j_{\epsilon(IC)} \over m_ec^2}
\sim \gamma \int_{\gamma} Qd\gamma \times \cases
{1/2 & if $\epsilon \ll \epsilon_{IC,KN}$ \cr
1 & if $\epsilon_{IC,KN} \ll \epsilon \ll \epsilon_{IC,s}$  \cr
qF_{KN} & if $\epsilon_{IC,s} \ll \epsilon \ll \epsilon_{IC,max}$ \cr },
\ee
where $\gamma \simeq \sqrt{(3/4)(\epsilon/\epsilon_{0,max})}$ if 
$\epsilon \ll \epsilon_{IC,KN}$, and $\epsilon \simeq \gamma$ if 
$\epsilon \gg \epsilon_{IC,KN}$; and
\be
 {4\pi \epsilon j_{\epsilon(syn)} \over m_ec^2}
\sim  \gamma \int_{\gamma} Qd\gamma \times \cases
{1/(2q) & if $\epsilon \ll \epsilon_{syn,KN}$ \cr
1/(2qF_{KN}) & if  $\epsilon_{syn,KN} \ll \epsilon \ll \epsilon_{syn,s}$  \cr
1/2 & if $\epsilon_{syn,s} \ll \epsilon \ll \epsilon_{syn,max}$ \cr },
\ee
where $\gamma=\sqrt{3\epsilon B_{cr}/(4B)}$.

For electron injection $Q \propto \gamma^{-p}$ and $p>1$, the 
various components are power laws, i.e., we have:
\be
{4\pi \epsilon j_{\epsilon(IC)} \over m_ec^2}
\propto \cases
{ \epsilon^{-p/2}  & if $\epsilon \ll \epsilon_{IC,KN}$ \cr
\epsilon^{-(p-1)}  & if $\epsilon_{IC,KN} \ll \epsilon \ll \epsilon_{IC,s}$ \cr
\epsilon^{-(p-1+\delta \alpha)} & if $\epsilon_{IC,s} \ll \epsilon 
\ll \epsilon_{IC,max}$ \cr }, \label{jIC2}
\ee
where 
$\delta \alpha \sim 1.5$ for 
$\alpha_0 < -1$, and  
$\delta \alpha \sim 1-\alpha_0$ for $0 <\alpha_0 < 1,$
while
\be {4\pi \epsilon j_{\epsilon(syn)} \over m_ec^2}
\propto \cases
{\epsilon^{-p/2}  & if $\epsilon \ll \epsilon_{syn,KN}$ \cr
 \epsilon^{-(p/2 + \Delta \alpha)} & 
if  $\epsilon_{syn,KN} \ll \epsilon \ll \epsilon_{syn,s}$  \cr
\epsilon^{-p/2} & if $\epsilon_{syn,s} 
\ll \epsilon \ll \epsilon_{syn,max}$ \cr },
\ee
where 
$\Delta \alpha=-(d\ln (1/F_{KN})/d\ln \epsilon)|_{b \propto \epsilon^{1/2}}$.
For $\alpha_0 < -1$, $\Delta \alpha \sim -0.75$, while
for $0 <\alpha_0 <1$, $\Delta \alpha \sim -(1-\alpha_0)/2$.
We should emphasize here, that estimation
of the hardening of synchrotron spectrum slope in the band 
$[\epsilon_{syn,KN}; \epsilon_{syn,s}]$ is very crude and corresponds
to its  maximum value, which as noted previously, can 
be reached  only for $q \gtrsim 10^3$. 
As shown in Fig.~\ref{fig8}, the synchrotron spectral 
hardening will be weaker for lower values of $q$.

Although  we have only presented  and discussed emission spectra computed
for an isotropic ambient radiation field, we would like 
to emphasize, that with our generic assumption of isotropic distribution of 
electrons, the electron energy losses, as well, as the
distribution of electrons (Eq. \ref{ng}) and their synchrotron emissivity 
(Eq. \ref{jsyn}), do not depend on angular distribution
of ambient radiation field.
Hence, in the case of a beamed ambient radiation field the IC spectrum can 
be simply computed from Eq. (\ref{jICt}) or  (\ref{jICtap}) by using there  
the electron distribution given by Eq. (\ref{ng}). In Fig.~\ref{fig10} we show 
such spectra for three different scattering angles and compare them with 
the one computed for the isotropic ambient radiation field.
One can see that the deeper one is in the KN regime, the weaker is
anisotropy of the scattered radiation. Suppression of the anisotropy is caused 
by the recoil effect.  

\subsection{Further effects that may modify the observed spectrum}

To highlight the spectral effects caused by 
the modification of the electron energy distribution due to Compton
scatterings in
the KN regime, we have considered
only spectra produced in the fast electron cooling regime,
and we have ignored possibly important 
processes such as photon-photon pair production 
and synchrotron self-Compton radiation. We also did not
consider effects due to the relativistic propagation of the 
source, which is important in objects like blazars. 
We discuss below  how  our results can be affected by 
the inclusion of some of these complications.
\smallskip

\noindent  
{\it The fast vs. slow cooling regime}

Assuming that electron
energy losses are dominated by radiative processes,
the cooling time scale for an electron of energy 
$\gamma$ is (see Eq. \ref{dotgtot})
\be t_c \equiv {\gamma \over |\dot \gamma|} =
{3 m_e c \over 4 \sigma_T u_B \gamma(1+qF_{KN})} \,. \label{tQ} \ee
For sources with a finite (comoving frame) lifetime $t_Q$, the only
electrons that have time to cool significantly are those with
$t_c < t_Q.$ Looking first at the Thomson limit ($\gamma
\lesssim \gamma_{KN}$) of this expression, one finds the usual
result that only electrons with $\gamma > \gamma_{c}$ have 
time to cool, where 
\be \gamma_c = {3 m_e c \over 4 \sigma_T u_0 t_Q} \,  \ee
for $q\gg 1.$  The number density of electrons at a given energy
that can accumulate in the source is roughly 
$Q(\gamma)\times \min[t_c(\gamma),t_Q].$  For power law electron injection,
the electron energy distribution of electrons therefore hardens
by $\Delta s=1$ for $\gamma < \gamma_c,$ leading to a hardening of 
the synchrotron or Compton spectrum  by $\Delta \alpha = 0.5.$

For $\gamma > \gamma_{KN},$ $q\gg 1,$ and a mono-energetic ambient photon
distribution, the 
the cooling time 
increases for $\gamma > \gamma_{KN},$  reaching a local maximum
at $\gamma\approx \gamma_s,$ when synchrotron cooling begins to 
dominate. Efficient KN cooling therefore requires 
$t_c(\gamma_s) <  t_Q.$ From Eq. (\ref{tQ}), 
we have
\be t_c(\gamma_s)= {3 m_e c \over 8 \sigma_T u_B \gamma_s}=
  {3 m_e c \epsilon_0  \over 2 \sigma_T u_0} {q \over b_s}
\, .\ee
Hence, fast cooling in the KN regime requires
\be {q \over b_s} < {2 \sigma_T u_0 t_Q \over 3\epsilon_0 m_e c} \, \ee
or, equivalently using $b_s \sim q^{2/3}$ for $1\ll q\lesssim 10^3$,
\be q < 
\left(2\sigma_T u_0 t_Q \over 3\epsilon_0 m_e c\right)^3 \, . \ee
Of course, if we have $b_{max} < b_s$, the upper limit on $q$
is correspondingly weaker. 

For a power law photon distribution with $0 \lesssim \alpha_0
\lesssim 1,$ we may use Eq. (\ref{tQ}) and approximation (6) to
show that $t_c \sim \gamma^{-\alpha_0}.$ For $\alpha_0>0,$
this is a monotonically decreasing function of $\gamma,$
so fast cooling throughout the KN regime is 
simply guaranteed
by the condition $t_c(\gamma_{KN}) \le t_Q,$ or equivalently
$\gamma_{KN} > \gamma_c.$  This can be translated into
the following requirement on the energy density of the external
radiation field,
\be
u_0 > {3 m_e c \over 4 \gamma_{KN} \sigma_T
t_Q} = {3 m_ec \epsilon_0 \over \sigma_T t_Q}\, .
\ee
For a photon distribution that is harder than
$\alpha_0=0$ or softer than $\alpha_0=1,$ the 
condition for a mono-energetic radiation field,
i.e., we require $t_c(\gamma_s) \le t_Q.$

\noindent
{\it Photon-photon pair production}

Because (a) the cross-section for photon-photon pair production
is similar in magnitude to that for Compton scattering, and (b)
the photon threshold energy for pair production, $\epsilon_+
\sim 1/\epsilon_{0,max},$ is almost the same as the electron
energy $\gamma_{KN} \sim 1/\epsilon_{0,max},$ it is often 
stated 
that strong pair production
is unavoidable in sources where KN effects are important.
This is not always true, however.

First, in many applications a better estimate for the 
threshold energy is in fact $\epsilon_+ = 2/\epsilon_{0,max}.$
Moreover, KN effects are actually important 
at energies well
below $1/\epsilon_{0,max},$ i.e., at 
$\gamma < \gamma_{KN} = 1/4\epsilon_{0,max}.$ Furthermore,
from Fig. 4., ${\cal A}(\gamma_{KN}) \sim 0.1 - 0.3,$ so
that an electron of energy $\gamma_{KN}$ Compton actually upscatters
photons to typical energies $\epsilon_{IC,KN} = {\cal A}_{KN}
\gamma_{KN} \ll \epsilon_+.$ Taking $\cal A$ at larger energies
to be $\sim 0.5,$ we see that we in fact need a source with $b_{max}>
b_+ = \gamma_+/\gamma_{KN} \sim \simeq 16.$
Since KN 
distortions of the electron spectrum already produce visible
distortions in the synchrotron spectrum for 
electron energies $b \sim 0.1,$ this means there is a factor
$\sim 100$ in electron energy for which KN corrections
are important but pair production is not possible. Since 
$\bar \epsilon_{syn} \propto b^2,$ this corresponds to 
a factor $10^4(!)$ in synchrotron frequency. 

Second, even if we have $b_{max} \gtrsim b_+,$ pair production
may still not be important. When one includes the effects of synchrotron
cooling, left out of KN pair production 
studies such as Zdziarski (1988), we have seen that one obtains a strong break
at $\epsilon_{IC,s}$ corresponding to the electron
energy $b_s$ where synchrotron cooling starts to dominate.
Thus, if $b_+<b_{max}$ but $b_+ > b_s,$ pair production
occurs but the luminosity of the pairs that are produced
(and the spectral distortions they induce) will not be bolometrically
important.
For a mono-energetic photon distribution, $b_s \simeq
q^{2/3},$ and one thus has $b_s < b_+$ for any $q\lesssim 60,$
{\it independent} of the actual value of the maximum electron energy, 
$b_{max},$ e.g., see Fig. 6a and Fig. 7. 
Note, though, that for 
an ambient photon
distribution that is not mono-energetic, e.g., a 
power law with $\alpha_0 \gtrsim -0.5,$
KN effects are not as strong as for the mono-energetic
case because lower energy photons that scatter in the Thomson regime
are available. 
The spectral break due to $\epsilon_{IC,s}$
therefore occurs at higher energy and pair production can be important for 
much lower $q,$ e.g., see Fig. 6b.

Even if pair production is energetically possible, the preceding
discussion says nothing about whether the optical depth to pair
production $\tau_{\gamma\gamma}$ actually exceeds unity, and it ignores
the effects of photon anisotropy (which raise the pair production
threshold). In a realistic source, the 
extent and geometry of the external radiation
field must be taken into as well as its absolute intensity. In galactic 
pulsar wind applications, for example, the ambient radiation field 
due to the companion star is often highly anisotropic in 
the source region of interest. In the most general case, 
$\tau_{\gamma\gamma}$ must be treated as a free parameter. In particular, if 
the source region has an effectively infinite lifetime and is not
expanding, which would induce adiabatic losses, we are essentially
free to choose as low an external field energy density as we like without
violating the fast KN cooling constraints discussed above. 
(If we choose
too low an external energy density, of course, higher order like
bremsstrahlung start to be important.) If the source is not 
static, though, there
are interesting limits we can place on $\tau_{\gamma\gamma}$
if we demand efficient KN cooling. For example,
consider a source of size $R$ with a characteristic lifetime
or expansion timescale $\sim R/c.$ For simplicity, assume
also that the radiation field is mono-energetic. Then, 
taking $\tau_{\gamma\gamma} \simeq 0.2 n_0\sigma_T R$ 
where $n_0=u_0/\epsilon_0$ and using the fast-cooling
condition of eqn. (35), 
we have 
\be
\tau_{\gamma\gamma} \gtrsim {3 \over 10} q^{1/3}.
\ee
This implies, for example, that we can be in the fast KN
regime and still have $\tau_{\gamma\gamma}<1$ for
$q \lesssim 37.$ 

\noindent
{\it Synchrotron-Self Compton (SSC) Effects}

In addition to the external ambient photons, synchrotron 
photons from the cooling electrons will always be present
in the source. The synchrotron photons spectrum is typically broad 
and extends to low energies, i.e., it is not well-approximated
by a mono-energetic photon distribution. The SSC spectrum produced by
synchrotron photon upscattering can thus be 
quite different in shape from the 
spectrum produced by the upscattering of external photons,
in particular it may extend to higher energies. This may be important
in certain applications. For a very weak external radiation field or
a very compact source, the Compton losses due the synchrotron
photons may in fact dominate over the losses due to the external
field, leading to significant changes in the equilibrium electron
distribution and thus the emergent spectrum. 
These will be explored further in a subsequent paper.
The analysis
is more complicated than in the present case because the
target radiation field and the electron distribution must
be determined simultaneously and self-consistently. (As we
have seen, many quantities depend sensitively on the details
of the low-energy photon spectrum.) 
Nonetheless,
the main effects described here still occur, e.g., the synchrotron
spectrum is harder than expected from a Thomson-limit analysis
and for sufficiently high electron injection energies, there
will be an energy $\gamma_s$ above which synchrotron losses
dominate, leading to a corresponding break in the Compton spectrum.

\noindent
{\it Relativistic source propagation effects}

If a source is moving with relativistic speed (bulk Lorentz factor 
$\Gamma  \gg 1$), and the external radiation field in such a 
frame is isotropic,
and with an energy density $u_0$ peaked at photon energies 
$\sim \epsilon_0$,
then in the source rest frame, the energy density of the external radiation is
$\Gamma^2$ times larger and the energies of seed photons are
$\Gamma$ times higher.
Then the rest frame quantities relevant to our discussion
of KN effects are
 $b= 4 \Gamma \gamma \epsilon_0$, and, in particular, 
$\gamma_{KN} \simeq  1/ (4 \Gamma \gamma \epsilon_0)$, and 
$q \simeq \Gamma^2 u_0/u_B$.
Since for $\Gamma \gg 1$  the head-on approximation applies, 
the IC Compton emissivity can be computed using Eq.(\ref{jICt}) with 
\be \cos \theta= -\cos \psi_{obs}^\prime =  - {\cos \psi_{obs}-\beta 
\over 1 -\beta\cos\psi_{obs}} \, ,
\ee 
where $\psi_{obs}^\prime$ is the angle between the jet axis and the direction
to the observer in the co-moving frame while $\psi_{obs}$ is 
the same angle but as measured in the lab frame. The observed source flux is 
then
\be \epsilon_{obs} F_{\epsilon_{obs}} = 
{{\cal D}^4 \int \epsilon j_{\epsilon}(\theta) dV \over d_L^2} \, \ee
where $\epsilon_{obs} = \epsilon {\cal D}/(1+z)$, 
${\cal D}=1/[\Gamma(1-\beta\cos\psi_{obs})]$ is the Doppler factor,
$d_L$ is the distance (luminosity distance for cosmological objects),
$z$ is the redshift, and $V$ is the volume of the source as measured
in its rest frame. The lifetime of the source measured in the 
lab frame is $\Gamma t_Q.$

\section{APPLICATIONS}

\subsection {Blazars}

The clearest and probably most numerous examples of 
high-$q$ (radiation dominated) non-thermal sources
are the powerful blazars where we think we
are seeing emission from a relativistic jet oriented towards us.
In many of these objects, the bolometric luminosity is strongly dominated by 
$\gamma$-rays (von Montigny et al. 1995; Mukherjee et al. 1997).  
A day-week variability time scales suggest that these sources
are located at (sub-)parsec  distances from the center.  There,
the external radiation field, as viewed  in the jet co-moving frame, 
is dominated by the powerful Broad Emission Line (BEL) region radiation.
For $u_0 \simeq L_{BEL}/(4 \pi r_{BEL}^2 c)$ and 
$u_B \simeq L_B/(\pi (r\theta_j)^2 c \Gamma^2)$, 
\be q(r \sim r_{BEL}) \simeq {\Gamma^2 u_0 \over u_B} \simeq 
25 {L_{BEL,45} (\Gamma/10)^2 (\Gamma \theta_j)^2 \over 4 L_{B,45} }     
\, , \label{qmax} \ee
where $L_B$ is the magnetic energy flux carried by the
jet and $\theta_j=R/r$ is the half-opening 
angle of a jet.

Let us determine now the  values of $\gamma_{KN}$ and $\gamma_s$
and of the corresponding IC and synchrotron photon energies for
such an external radiation field, assuming that 
$q \gg 1$ and  $\gamma_{max} > \gamma_s$.
Noting that the energies  of broad emission lines 
peak around $\sim 10$eV ($\epsilon_0 \approx 2\times 10^{-5}$)
and that they are seen 
in the jet co-moving frame as boosted by a factor
$\sim \Gamma$, i.e.
$b=4 \epsilon_0 \gamma \Gamma$, we have
\be \gamma_{KN} = {1 \over 4\epsilon_0\Gamma} \sim  10^3 
(\Gamma/10)^{-1} \, \ee
and 
\be \gamma_s = b_s \gamma_{KN} \sim 
  10^4 (q/30)^{2/3} (\Gamma/10)^{-1} \, \ee
These electrons Comptonize external photons up to average energies 
\be \epsilon_{IC,KN}^{obs} \simeq {\cal A}_{KN}  \gamma_{KN} {\cal D} 
\sim 2 \times 10^3 ({\cal A}_{KN}/0.14) ({\cal D}/\Gamma) \, \,  \, 
(\sim 1 {\rm GeV}\,  ...) , \ee
and
\be \epsilon_{IC,s}^{obs} \simeq {\cal A}_{s}  \gamma_s {\cal D} \simeq
6 \times 10^4 ({\cal A}_{s}/0.5)(q/30)^{2/3} ({\cal D}/\Gamma)  \, \, \,
(\sim 30 {\rm GeV}\,  ...) , \label{eICs} \ee
and produce synchrotron photons with average energies  

\be \epsilon_{syn,KN}^{obs} = {4\over 3} \gamma_{KN}^2 {B \over B_{cr}}
{\cal D} \sim 2 \times 10^{-7} 
{L_{B,45}^{1/2} ({\cal D}/\Gamma) \over R_{BEL,18} (\Gamma/10)} \, \, 
(\sim 3 \times 10^{13}{\rm Hz} ...) \, , \ee
\be \epsilon_{syn,s}^{obs} = b_s^2 \epsilon_{syn,KN}^{obs} \simeq
q^{4/3} \epsilon_{syn,KN}^{obs} 
\sim 2 \times 10^{-5} (q/30)^{4/3}  
{L_{B,45}^{1/2} ({\cal D}/\Gamma)
\over R_{BEL,18} (\Gamma/10)} 
(\sim 3 \times 10^{15}{\rm Hz} ...) \, . \ee
Hence, sources with $q \gg 1$, $1 < b_{max} < b_s$ and  power-law electron
injection should produce  synchrotron ``bumps'' peaking in the IR-UV 
spectral band, with the closer a given $b_{max}$ is to $b_s,$ 
the more prominent the bump. As discussed, the presence of ``excess'
synchrotron emission (the bump) in this case 
is simply due to KN effects and should not 
be interpreted as indicating the presence of a new electron acceleration
component or a hardening of the low energy injection spectrum.

We have just estimated the electron injection parameters that
would put a luminous blazar into the KN regime studied here.
Are there any blazars that actually populate this region of 
parameter space?
Since for even the brightest blazars, EGRET could not detect
gamma-rays much beyond 
$1$ GeV, the extension of the electron 
energy distribution into KN regime cannot be established using 
EGRET data. However, quite significant constraints on the high energy tails 
of the electron energy distribution  are provided by observations of 
their synchrotron spectra. In the study of Padovani et al. 2003, 
about half of the objects in their sample of powerful blazar
objects have  
synchrotron spectra peaking at $\nu > 3 \times 10^{13}$Hz,
which implies $b>1$. 
In most cases the
spectra  steepen 
in the UV band, indicating an
injection function with a cutoff or break at $b \sim$ a few,
i.e., KN effects may be moderately important. There
are few of these blazars, though, that have  
a synchrotron peak clearly located in the UV to soft X-ray band
and may have $b_{max} \gg 1.$ 
For these objects, assuming the
external BEL radiation field is dominant with the characteristics
described above,
we then predict
a high energy spectral break of the IC component at 
$\epsilon_{IC,s} \sim 30$ GeV, independent of the 
observed synchrotron cutoff energy
(i.e., $b_{max}$) provided that it is sufficiently large. 
If the electron injection spectrum for these objects is not
unusually soft, i.e., we have $p \lesssim 3,$ 
then as shown in Fig. 6-8, for example,
a significant of their bolometric luminosity actually emerges
in the synchrotron component. In particular, their 
IC luminosity should be comparable
to their synchrotron luminosity, i.e., they would {\it not}
be gamma-ray loud objects (with $L_{IC}/L_{syn}\gg 1$).

It would be very convenient if powerful blazars had gamma-ray
spectra extending to TeV energies, e.g., they would provide 
very bright sources that could used to constrain the intensity of 
the extragalactic background light via the absorption 
of their gamma-rays (Coppi \& Aharonian 1999).
Unfortunately, for the BEL parameters
we have used, this is impossible unless $q$ is extremely
(and implausibly) large. One caveat to this conclusion is
that we have not included SSC effects in our estimates, and
because the synchroton emission is much broader in energy than the BEL
one, $\gamma_s$ and $\epsilon_{IC,s}$ 
could move to higher values.
Note, though, that $\epsilon_+,$ 
the energy above which pair production on the BEL becomes
possible, is only $\simeq 50$ GeV. Strong Compton TeV emission therefore
seems unlikely unless it occurs far from the BEL and the typical
ambient photon energy is in the near-infrared range.

\subsection{Micro-blazars?}

Only two or three EGRET  sources have been identified with
micro-quasars (Paredes et al. 2000; Massi et al. 2004; 
Combi et al. 2004). The fact that these objects are 
High Mass X-Ray Binary (HMXB) systems containing  massive  
and very luminous companion stars strongly suggests 
a Compton origin for their $\gamma$-rays.
As in blazars, the EGRET observations unfortunately do not provide constraints 
 on $\gamma_{max}$ for these objects. Furthermore, because these sources are
relatively weak and completely dominated  
in the optical/UV band by radiation from the companion stars,
their synchrotron component cannot be identified.
However, GRS 1915+105 proves that XRB have the ability 
to produce much more powerful 
and relativistic jets (Fender \& Belloni 2004) than in the sources
just mentioned. If such a powerful jet were to occur
in an HMXB  and it pointed toward us, 
we would see a {\it micro-blazar}, with relativistically boosted 
non-thermal radiation dominating over all spectral bands
(Georganopoulos, Aharonian, \& Kirk 2002). Since the spectra of HMXB
companion stars peak at $\sim 10$ eV, the same value for the BEL in 
quasars, the values of
$\gamma_{KN}$ and $\gamma_s$, and of $\epsilon_{IC,KN}$ and
$\epsilon_{IC,s}$ are of the same order as for blazars and, therefore, 
like blazars, they are expected to be GeV emitters
and not strong TeV emitters.
Due to the much stronger magnetic fields in micro-quasar jets, the 
synchrotron spectral bumps, 
produced if $b_{max} \gg 1,$ 
are predicted to peak in the UV/soft X-ray band.
Hence, if some of the ULX (Ultra-Luminous X-ray) sources are
in fact micro-blazars pointed at us, 
they should be strong $\gamma$-ray emitters (Georganopoulos {\it et al.} 2002).
 
However, we must remember, that very large $q$ is available only
on size scales comparable to those of the binary system. For electron
acceleration occurring well down the jet, outside the binary system, 
the companion star radiation chases the relativistically moving emission
region from behind and its energy density is thus Doppler deboosted
when
viewed in the jet frame. (The energy density of the companion star radiation
field will be further reduced by the usual factor, $\propto 1/r^2$ 
where $r$ is the distance from the binary, but this
effect can be canceled out by the fact that the magnetic field energy  in a 
conically expanding jet also drops as $1/r^2.$) 
Closer in to the central object, jets typically have much 
stronger magnetic fields and, therefore, synchrotron radiation
will dominate electron energy losses,  even in the Thomson regime. 
In this case the IC spectrum is expected to break at $\epsilon_{IC,KN}$, i.e., 
$\sim$ 1 GeV. This would be the case if, as in blazars, 
the jet energy  is dissipated  at $10^{3-4} (r/r_g,)$ 
where $r_g$ is the 
gravitational radius of the compact object, i.e., on scales 
$10^3$ times smaller than  the typical size of an XRB system.

It should be noted that because the 
radiation field of the companion star 
is not symmetric about the jet axis,
 detailed computations of the non-thermal spectra
from XRB require the integration of emissivities given by Eq. (\ref{jICt}) 
over the energy distribution of the external radiation field, 
taking into account 
that in the source comoving frame, 
the radiation is boosted by a Doppler factor 
that depends on  the direction of the incoming photons 
(Khangulian \& Aharonian 2005). In other words, the
exact geometry of the system is important.

\subsection{Pulsars in HMXB}

PSR B1259-63  and PSR J0045-73  provide examples of 
non-accreting binary pulsars with massive companions
(Johnston et al. 1992; Kaspi et al. 1994);
and  PSR B1259-63 was recently identified as a TeV source 
(Aharonian et al. 2005).
Ball and Kirk (2000) envisaged two scenarios for production
of high energy radiation in such systems, Comptonization of  the
radiation field of the companion luminous star
directly by the pulsar ultra-relativistic wind (i.e.,
a bulk-Compton scenario)
or 
by particles accelerated 
in the terminal wind shock formed due to the 
confinement of the pulsar wind by the 
wind from the companion star.  Since for finite values of $q\gg1,$ 
the IC spectrum has a high energy break
at $\epsilon_{IC,br}=\min(\epsilon_{IC,s},
\epsilon_{IC,max}),$
the condition for efficient
TeV production is $\epsilon_{IC,s} \ge 10^6$
and $\epsilon_{IC,max}  \ge 10^6.$
Since 
$\epsilon_{IC,s} \sim \gamma_s b_s \gamma_{KN}$, 
where $\gamma_{KN} = 1/(4\epsilon_0) \sim 10^4$, this translates
into the condition 
$b_s  > 100$, i.e., for a mono-energetic field, 
we again need a very large $q \gtrsim 10^3$.
In the shocked wind scenario, 
this condition provides an upper limit 
on the strength of the magnetic field in the shocked plasma.
In the bulk-Compton scenario, the condition is satisfied even
for strongly magnetized winds. This is because electrons are cold
and frozen to the magnetic field lines, hence their energy losses
via  the synchrotron mechanism are negligible and, effectively, the value
of $q$ is infinite. Of course, to reach  TeV energies in this scenario, 
the bulk Lorentz factor of the wind is required to be of the order $10^6$,
which is consistent with estimations of the wind speed in the Crab Nebula
(Rees \& Gunn 1974).

\subsection{Kiloparsec scale jets}

Jets in quasars encounter a variety of radiation fields as 
they make their way out from the central black hole, 
starting from the radiation field of the black hole accretion disk
and ending with
the cosmic microwave background (CMB). 
For jets that are relativistic from the start, the largest $q$ would be 
reached  near the base of the jet.
The lack of  bulk Compton features in 
the soft X-ray band suggests this is not the case (Moderski et al. 2004),
and thus that the energetically dominant Compton interactions of a jet with
external radiation field seem to take 
place not earlier than in  the BEL region, i.e. around $0.1-1$ parsec from 
the center. 
Large values of $q$ are also expected
in sources triggered at 1-10 parsec distances where the
diffuse component of the external field is likely dominated
by the thermal emission from hot dust (Sikora et al. 2002).
At progressively larger distances, the jet undergoes Compton interactions 
with narrow emission 
lines, with stellar radiation, and finally the CMB. 
As of now there is no direct evidence that at such distances 
electrons are injected with $b_{max} \gg 1$, but this may simply
be due to the
sensitivity and angular resolution of present $\gamma$-ray detectors.
Indirect evidence in favor of $b_{max} \gg 1$ and $q\gg 1$ for
kiloparsec scale jets may come from the work of  Dermer and Atoyan  (2002).
They argued that a purely synchrotron emission model, with a 
jet magnetic energy density less than the CMB energy density, 
can successfully
explain not just
the observed radio-to-optical jet radiation,
but also the X-ray flux detected by Chandra, despite the fact that 
the Chandra flux
often lies above an extrapolation of the radio-optical spectrum. This is 
achieved by the formation of a synchrotron bump above $\epsilon_{syn,KN}$
due to the KN effects we have described here.
We note, however, this 
particular model requires extremely large $\gamma_{max}$, at least
one order of magnitude larger 
than $\gamma_{KN} = 1/( 4 \Gamma \epsilon_{CMB}) \simeq
2 \times 10^7/(\Gamma/10)(1+z)$, where $\epsilon_{CMB}(z=0) \sim 10^{-9}$.

\section{CONCLUSIONS} 

We have studied, both analytically and using accurate numerical codes,
the electromagnetic spectra produced by
relativistic electrons in a magnetized
non-thermal source that is immersed in a dense radiation field originating
outside the source. 
We consider the case when the steady-state electron energy distribution is 
determined by the injection energy spectrum of the accelerated electrons and
their energy losses, dominated by synchrotron radiation and 
Compton scattering which may extend deeply into the KN regime.
We concentrate on the poorly studied
region of parameter space in which 
the energy density of the radiation
field inside the source exceeds that of the magnetic field, i.e., 
values of the parameter $q=u_0/u_B > 1.$ 
Fig. 10 summarizes the three main regions we find
for the overall parameter space for our problem:

\noindent
-- In zone I ($q<1$), the electron distribution is determined
by synchrotron cooling and KN effects do not appear in the synchrotron
spectrum. The Compton spectrum, however, shows a strong break
at $\epsilon_{IC,KN}$ due to the strong KN reduction in the
Compton scattering rate that starts for electron energies 
$\gamma > \gamma_{KN}.$ Assuming the maximum electron acceleration energy,
$\gamma_{max},$ is sufficiently large that $\epsilon_{IC,max}
> \epsilon_{IC,KN},$ the position of this break is {\it independent}
of $\gamma_{max}.$ In zone Ib ($1<q\lesssim 3$), we start to
see a hardening 
of the electron distribution that is reflected in the synchrotron emission
spectrum. The hardening occurs because Compton cooling is now 
an important contribution to the total electron cooling rate,
and Compton losses decrease at high energies due to the KN effect.
The effect is not large, however. The Compton spectrum does not show
significant differences because synchrotron losses start to 
dominate again at $\gamma_s < \gamma_{KN},$.

\noindent
-- In zones II and III,
$q \gg 1,$ and the distortion in the electron distribution
due to the reduction in Compton cooling is very large. The synchrotron
spectrum correspondingly shows a strong, hard excess over the
Thomson limit asymptote. The Compton spectrum below $\epsilon_{IC,s},$
however, does not show a strong deviation from the low energy
Thomson limit because the KN decline in the Compton rate is 
compensated by the corresponding increase in the electron density.
The distinction between zone II and III is that in zone II, 
the maximum electron energy is such that $\gamma_{max} > \gamma_s.$
Above $\gamma_s,$ synchrotron cooling dominates.   The energy distributions
of the electrons and the synchrotron radiation reach asymptotes
with the same slopes as in the Thomson regime but with 
amplitudes enhanced by a factor $q.$ The Compton spectrum, on 
the other hand, shows a 
strong break at $\epsilon_{IC,s}$ because the KN reduction
in the scattering is no longer compensated by a hardening of 
the electron energy distribution.
In zone II, then, the Compton spectrum breaks
at an energy {\it independent} of $\gamma_{max}.$ 
In zone III, instead,
the synchrotron losses never dominate, and 
the location of Compton high energy break is determined by the 
maximum electron energy just as it is for the synchrotron component.
The combination of the high energy break and the hardening
of the synchrotron spectrum at lower energies 
leads to the formation of a strong synchrotron bump.

The specific conclusions of our study are:

\noindent
$\bullet$
The IC spectra have high energy breaks 
at $\epsilon_{IC,max}$,
if $b_{max} < b_s$, or at $\epsilon_{IC,s}$, if $b_{max}> b_s$.
The former is related to the high energy cut-off of electron injection 
function, the latter to the strong  steepening of the IC spectrum,
caused by domination of energy losses of electrons with $\gamma > \gamma_s$
by synchrotron mechanism; 

\noindent
$\bullet$
Synchrotron spectra undergo strong hardening at $\epsilon > \epsilon_{syn,KN}$,
with $|\Delta \alpha|$ reaching
$\sim 0.5-.75$ for $q > 30$. Hence, for very hard electron  
injection spectra, with $p < 1$, the spectral index $\alpha$ ($=0.5 -
|\Delta \alpha|$) can even reach 
negative values. 
The hardening 
is visible already at $\epsilon \simeq 0.1 \epsilon_{syn,KN}$;

\noindent
$\bullet$
For $1 < b_{max} \le b_s$, the hardening of the synchrotron spectrum 
combined with the high energy break at $\epsilon_{s,max}$ leads to the
formation a ``bump'' in the high-energy portion of the synchrotron spectrum.
For $b_{max} \gg  b_s$, the hardening of the synchrotron spectrum stops
at $\epsilon_{syn,s}$ and the spectrum continues with the same slope as in
the Thomson regime but with a normalization $q$ times larger;

\noindent
$\bullet$  For hard electron injection functions ($p<2$) and 
$b_{max} > b_s$, the luminosity of the synchrotron component
is larger than luminosity of the IC component, even for $q \gg 1$.
This is because for a hard injection function  most of the power 
is supplied to 
electrons with $\gamma > \gamma_s,$ and the energy losses for these electrons
are dominated by synchrotron radiation. For $p=2,$ the  
luminosities 
of the synchrotron and IC spectral peaks are of the same order; 

\noindent
$\bullet$ When KN effects are important, both the IC and
especially the synchrotron component can have spectra {\it harder}
than the hardest spectrum possible in the Thomson limit for fast cooling
electrons ($\alpha=0.5$); 

\noindent 
$\bullet$ Generically, photon-photon pair production of the 
Compton gamma-rays on the ambient radiation field may be important
for $\gamma_{max} \gg \gamma_{KN}.$ However, for a pair production
energy threshold exceeding $\epsilon_{IC,s}$, the fraction of the $\gamma$-ray
luminosity converted into pairs is not significant,
even if the opacity for pair production is large;

\noindent
$\bullet$ 
The continuous energy loss approximation for the evolution of the 
electron distribution appears to work reasonably well, even
for $\gamma_{max} \gg \gamma_{KN}.$ Use of this approximation
can save considerable computing time;

\noindent
$\bullet$  The KN effects we have discussed 
can be important in powerful blazars and  HMXB, with 
the latter including accreting compacts objects and rotationally
powered pulsars.

\acknowledgments

This project was partially supported by Polish KBN grants 
1 P03D 00928, PBZ-KBN-054/P03/2001.
M.S. and P.C. thank the Fellows of the MPI f\"ur Kernphysik, 
and SLAC and KIPAC for
their hospitality and support during their stays there.
P.C. was also supported in part  by 
a Yale University Senior Faculty Fellowship.

\appendix

\section{APPENDIX}

\subsection{Scattering of directed photon beams
on isotropically
distributed relativistic electrons}

The general formula for the distribution in energy and angle of
the scattered photons per electron  per unit time for the photon beams
is given by Eq. (14) in Aharonian \& Atoyan (1981). For 
$\epsilon \gg \epsilon_0$ and $\gamma \gg 1$, this formula takes the form 
(see Eqs.[20] and [21] in Aharonian \& Atoyan 1981)
\be {\partial \dot N_{sc}(\epsilon, \gamma, \theta) \over
\partial \epsilon \partial \Omega} \simeq {3 \over 16 \pi} c\sigma_T 
\int_{\epsilon_{0,m}(\gamma,\theta)}
 {n_{\epsilon_0} \over \epsilon_0 \gamma^2} 
f(\epsilon, \epsilon_0, \gamma, \theta) d\epsilon_0 \, , \label{Nsctheta} \ee
where  $\epsilon_0$ and $\epsilon$ are energies of the incident and
scattered photons in $m_ec^2$ units, respectively, $\theta$ is
the scattering angle, $n_{\epsilon_0}$ is the photon number density
per energy, 
\be
\epsilon_{0,m}(\gamma,\theta) = {\epsilon \over 
2(1-\cos\theta) \gamma^2 (1-(\epsilon/\gamma))} \, , \label{epsmt} \ee
and
\be f(\epsilon, \epsilon_0, \gamma, \theta)
= 1 + {w^2 \over 2(1-w)} - {2w \over b_{\theta}(1-w)} +
{2w^2 \over b_{\theta}^2(1-w)^2} \, \label{ftheta} \ee
where
$b_{\theta}= 2(1-\cos\theta)\epsilon_0\gamma$, and $w=\epsilon/\gamma$.

\subsection{Scattering of isotropically
distributed photons  on isotropically
distributed relativistic electrons}

For an isotropic radiation field
\be {\partial \dot N_{sc}(\epsilon, \gamma) \over
\partial \epsilon \partial \Omega} = {1\over 4\pi} 
{\partial \dot N_{sc}(\epsilon, \gamma) \over \partial \epsilon} =
{3 \over 16 \pi} c\sigma_T 
\int_{\epsilon_{0,m}(\gamma)}
{n_{\epsilon_0} \over \epsilon_0 \gamma^2} 
f_{iso}(\epsilon, \epsilon_0, \gamma) d\epsilon_0 \, , \label{Nsc} \ee 
where (see Eq.[23] in Aharonian \& Atoyan 1981)
\be
\epsilon_{0,m}(\gamma) = {\epsilon \over 4 \gamma^2 (1-(\epsilon/\gamma))}
\, , \label{epsm} \ee
and 
\be f_{iso}(\epsilon, \epsilon_0, \gamma)= 
{1 \over 4\pi} \int_{\theta_{min}}
f(\epsilon, \epsilon_0, \gamma, \theta) d\Omega 
= {1\over 2 } \int^{\cos \theta_{min}}
f(\epsilon, \epsilon_0, \gamma, \theta) d\cos \theta \, , \label{fis} \ee
where $\cos\theta_{min}=1-2w/(b(1-w))$. Integration in Eq.~(\ref{fis}) 
can be performed analytically giving 
(see Eq.[22] in Aharonian \& Atoyan 1981) 
\be f_{iso}(\epsilon, \epsilon_0, \gamma)
= 1+{w^2 \over 2(1-w)} + {w \over \tilde b(1-w)} -
{2w^2 \over {\tilde b}^2(1-w)^2} - {w^3 \over 2{\tilde b}(1-w)^2} - 
{2w \over {\tilde b}(1-w)} \ln{{\tilde b}(1-w)\over w} \, , \label{fiso} \ee
where 
$\tilde b=4\epsilon_0 \gamma$ (note that $f_{iso}$ is fully  equivalent to the
term bracketed  in Eq.[9] in Jones [1968]). 

\subsection{Electron energy losses}

The rate of inverse-Compton energy losses of electrons is
\be |\dot \gamma|_{IC} 
\simeq
{3\over 4} c \sigma_T {1 \over \gamma^2} \int {n_{\epsilon_0} \over \epsilon_0}
\left[ \int f_{iso}(\epsilon, \epsilon_0, \gamma) \epsilon d\epsilon \right]
 d\epsilon_0 \, . \label{dotg1} \ee
The inner integral has an analytical solution (see Eq. [46] in Jones 1968) 
\be \int f_{iso}(\epsilon, \epsilon_0, \gamma) \epsilon d\epsilon=
{\gamma^2 g(\tilde b) \over \tilde b} \, \label{Intfiso} \ee
where
\be g(\tilde b)= \left( {1 \over 2} \tilde b + 6 + {6 \over \tilde b} \right) 
\ln (1+{\tilde b}) -
\left( {11 \over 12} \tilde b^3 +  
6 {\tilde b}^2 + 9 \tilde b + 4 \right) {1 \over (1+\tilde b)^2}
- 2 + 2 {\rm Li}_2 (-\tilde b)  \, . \label{gb}
\ee
and $Li_2$ is the dilogarithm.
Hence,
\be  |\dot \gamma|_{IC}  =
 {4c \sigma_T \over 3 }\, \gamma^2 \,
\int f_{KN}(\tilde b)\epsilon_0 n_{\epsilon_0} d \epsilon_0 \, \label{dotg2}
\ee
where 
\be f_{KN}(\tilde b)= 9 g(\tilde b)/{\tilde b}^3 \, . \label{fKN} \ee

\clearpage

\begin{figure}
\centering
\includegraphics[width=5in,angle=0]{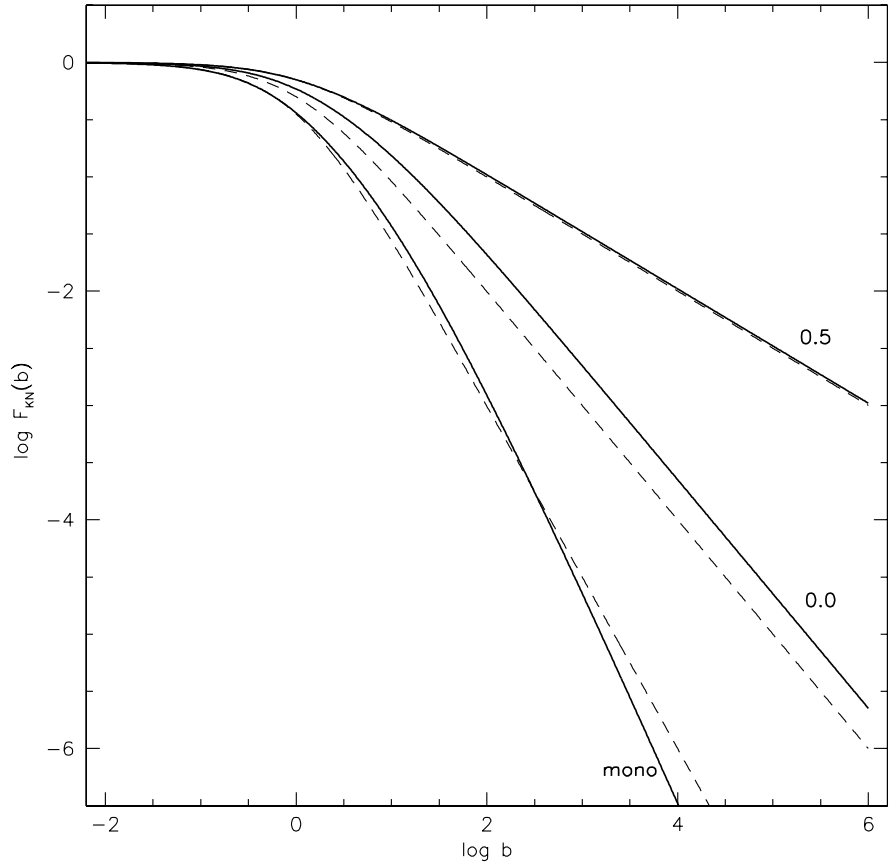}
\caption{The function $F_{KN}(b)$ computed for  mono-energetic 
(``mono'') and
power law ($\alpha_0=0.0$ and 
$\alpha_0=0.5$) 
energy distributions of the external photon field. The solid
lines show the results of the exact calculations while the dashed lines
are the analytical approximations.}
\label{fig1}
\end{figure}

\clearpage

\begin{figure}
\centering
\includegraphics[width=5in,angle=0]{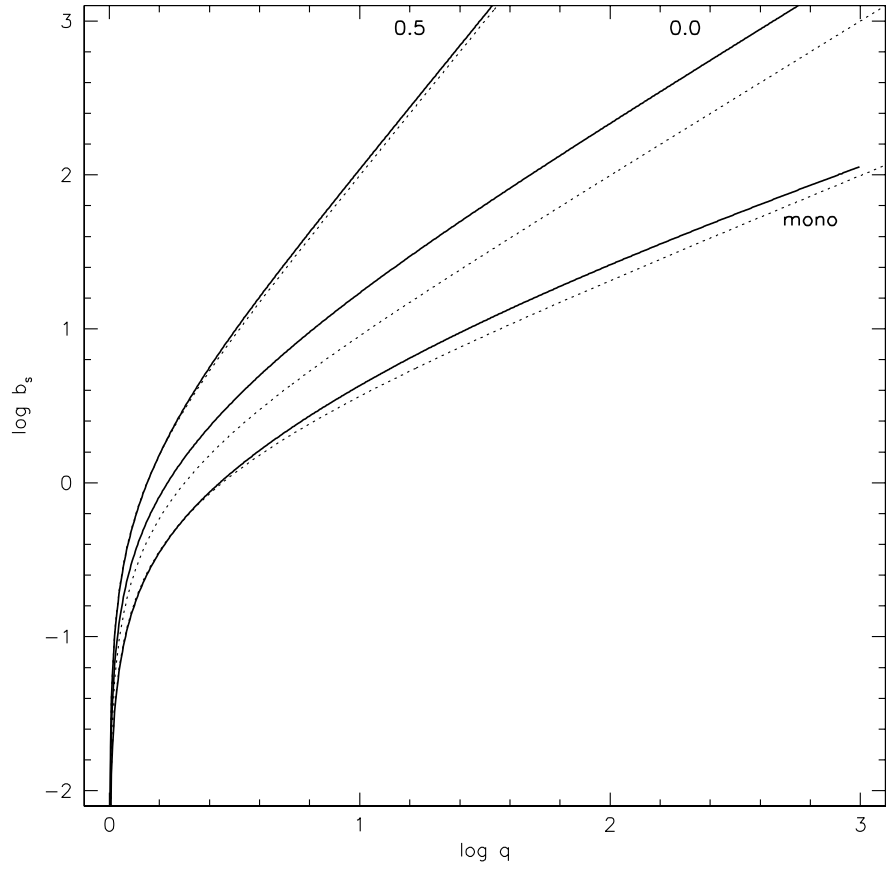}
\caption{ $b_s$ vs $q$ for: mono-energetic and power law 
ambient radiation fields. 
Solid lines -- exact results, dashed lines -- analytical  aproximations.}
\label{fig2}
\end{figure}

\clearpage

\begin{figure}
\centering
\includegraphics[width=5in,angle=0]{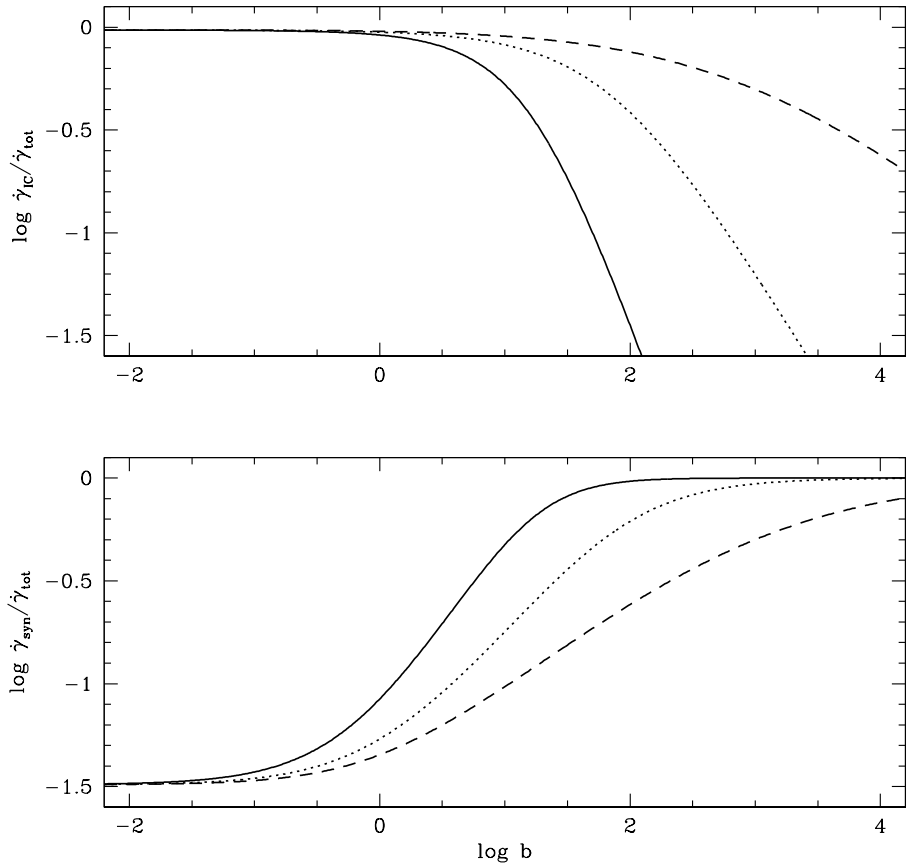}
\caption{ The relative IC and  synchrotron energy losses for
an ambient radiation energy distribution that is mono-energetic
(solid lines) or  power law ($\alpha_0=0.0$ -- dotted lines;
$\alpha_0=0.5$ -- dashed lines).}
\label{fig3}
\end{figure}

\clearpage

\begin{figure}
\centering
\includegraphics[width=5in,angle=0]{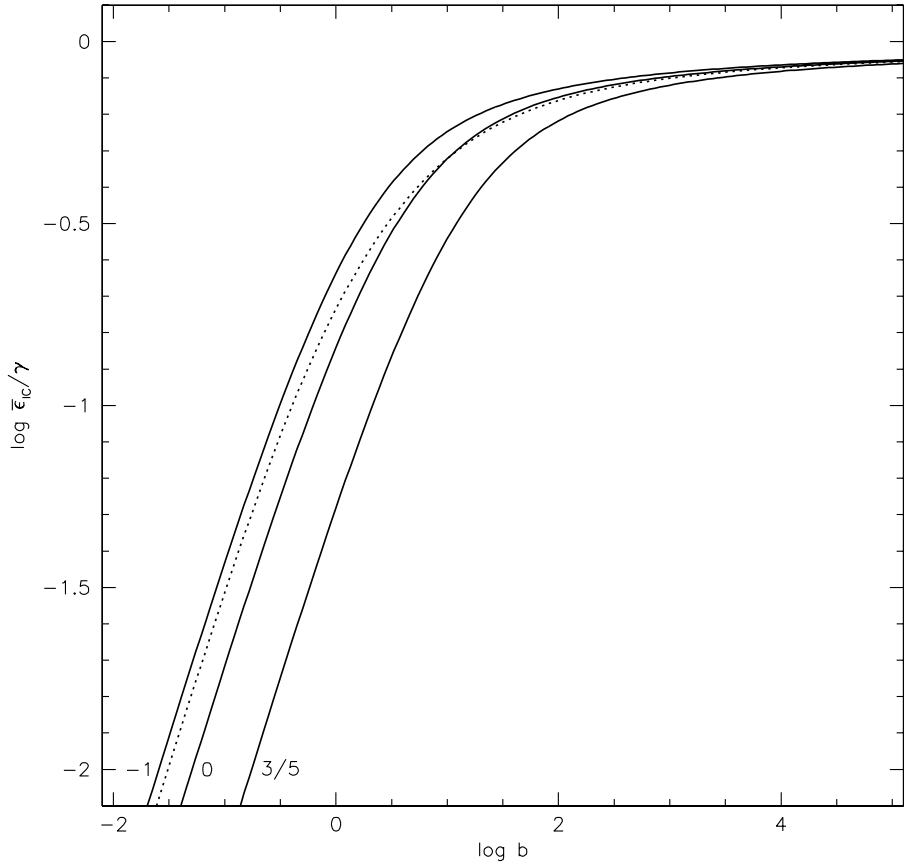}
\caption{Inelasticity, ${\cal A}=\bar \epsilon_{IC}(\gamma)/\gamma$ 
for electron Compton scatterings off a  
mono-energetic ambient radiation field. 
Dotted line -- isotropic ambient 
radiation field. Solid lines -- beamed ambient radiation field,
for scattering angles $\cos \theta = -1.0, 0.0, $and$ 0.6.$}
\label{fig4}
\end{figure}

\clearpage

\begin{figure}
\centering
\includegraphics[width=5in,angle=0]{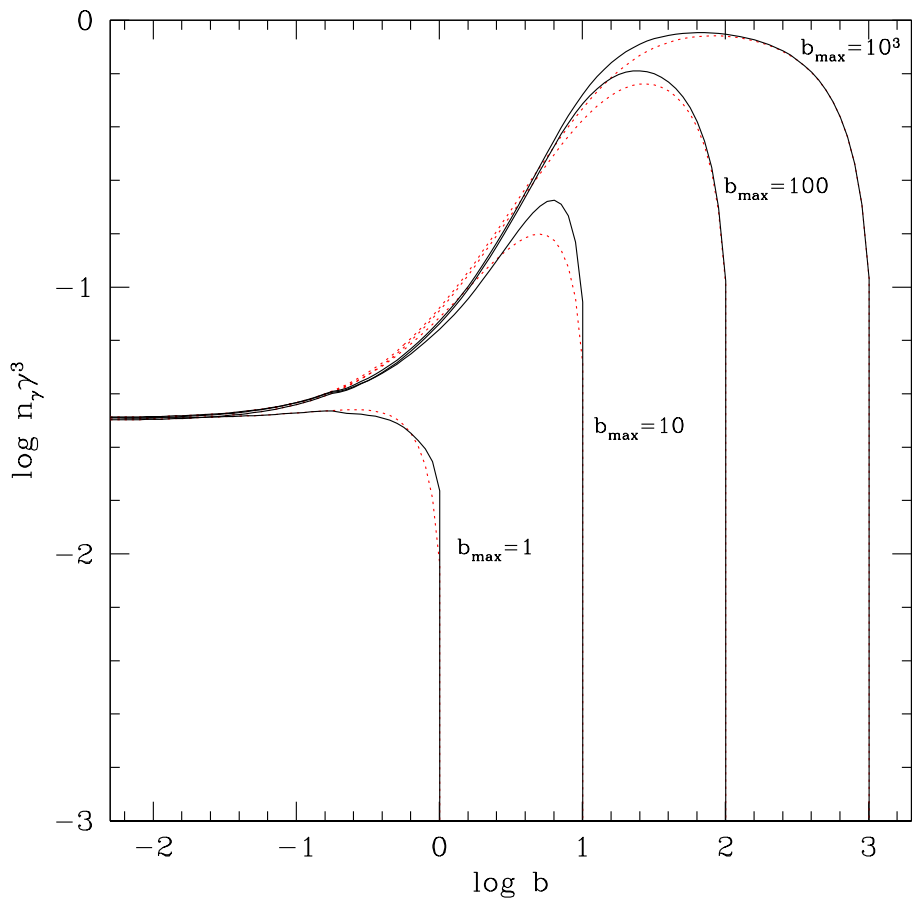}
\caption{ Steady state electron energy distributions  for power-law 
electron injection function, $Q \propto \gamma^{-p}$, and mono-energetic
ambient radiation field.  Solid lines --
exact results; dotted lines --  results obtained using the 
continuous energy loss approximation for all Compton scattering.
The model parameters are: $p=2$, $q=30$; $b_{max}= 1; 10; 10^2; 10^3$. }
\label{fig5}
\end{figure}

\clearpage

\begin{figure}
\centering
\includegraphics[width=5in,angle=0]{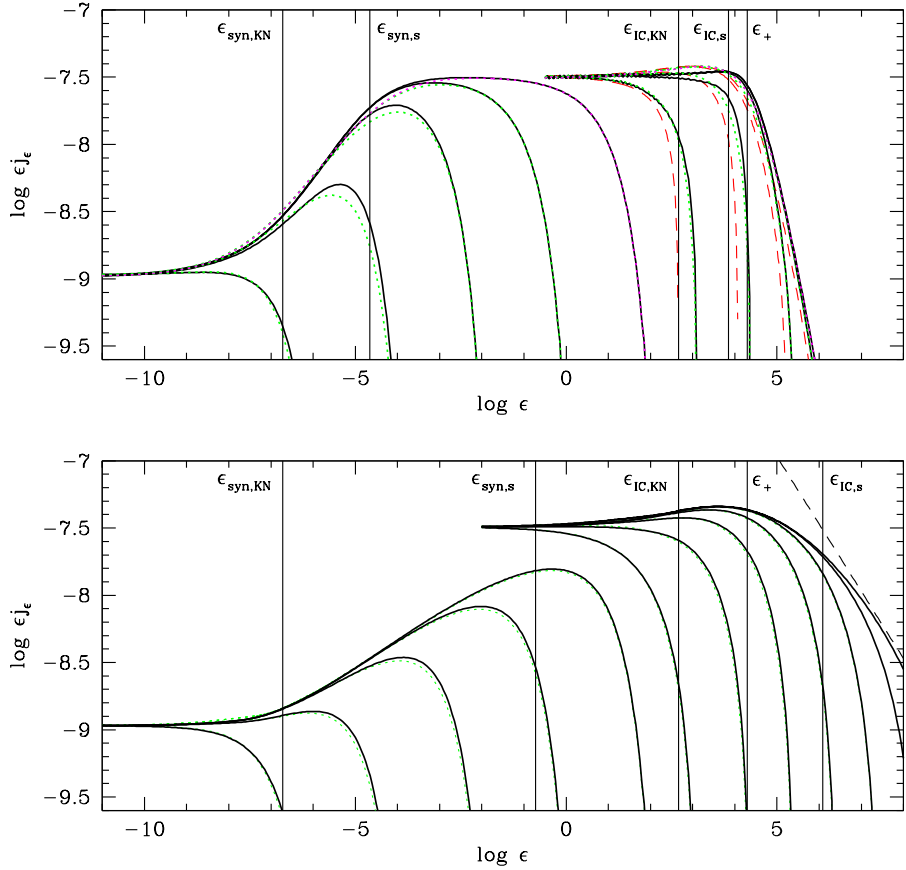}
\caption{Inverse-Compton plus synchrotron spectra of  steady sources.
with model parameters $p=2$; $q=30$; $b_{max}=1, 10, 10^2, 10^3, 10^4;$
$\epsilon_0=10^{-4}$; $B=1$Gauss.
(a) Upper panel -- mono-energetic ambient radiation field.
(b) Lower panel -- power-law ambient radiation field with
$\alpha_0 = 0.5.$ Solid lines -- exact calculations. Dotted lines -- 
calculations using continuous energy loss approximation.
Dashed lines -- Compton spectra computed using the
continuous energy loss approximation and the delta-function
approximation. The dot-dashed line in the lower panel is the
asymptotic power law ($\alpha=-0.5$) for the IC spectrum at $\epsilon > 
\epsilon_{IC,s}$ given by Eq. (\ref{jIC2}). To show convergence to this
spectrum for increasing $b_{max},$ the lower panel also shows the IC 
spectra obtained for $b_{max}=10^5, 10^6,$ and $10^7.$ }
\label{fig6}
\end{figure}

\clearpage

\begin{figure}
\centering
\includegraphics[width=5in,angle=0]{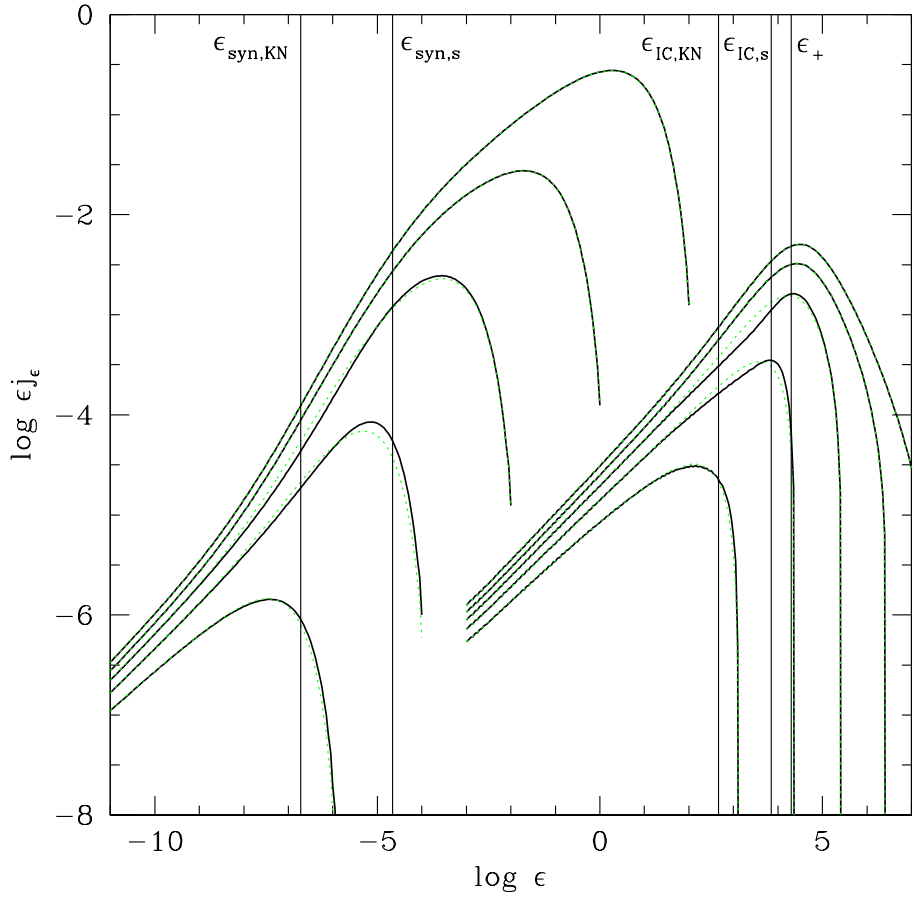}
\caption{Same as the upper panel of Fig.~\ref{fig6} (mono-energetic
radiation field), but for $p=1$.}
\label{fig7}
\end{figure}

\clearpage

\begin{figure}
\centering
\includegraphics[width=5in,angle=0]{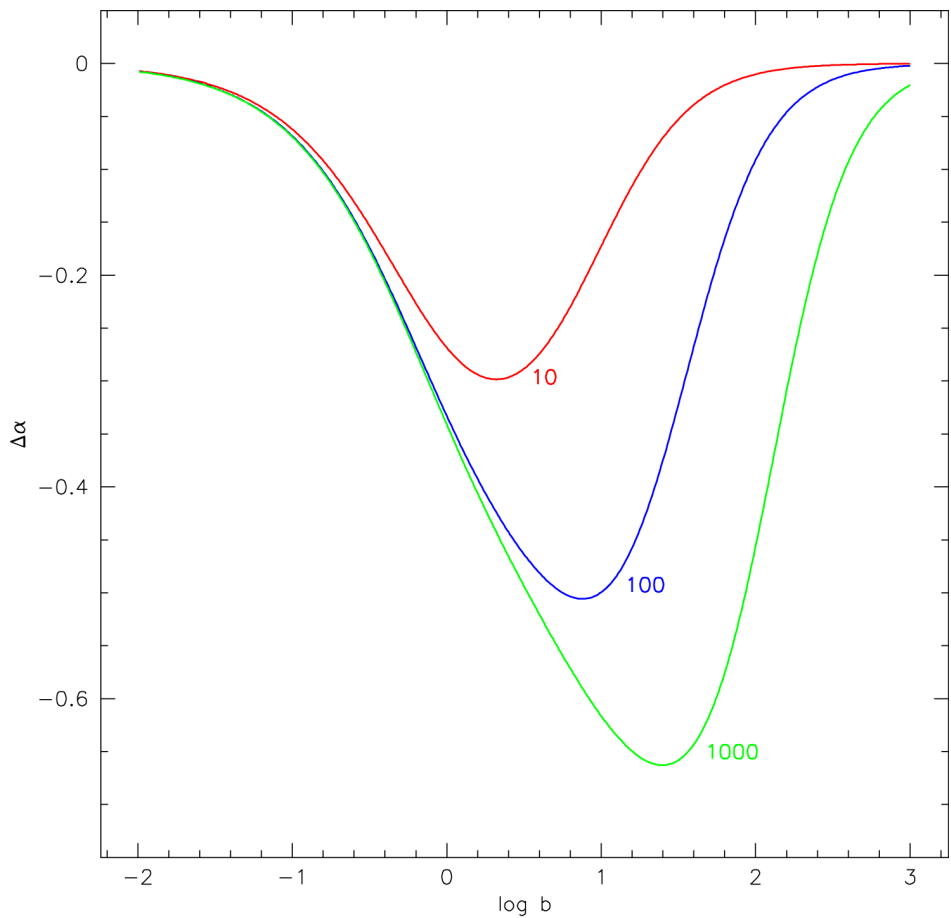}
\caption{ Synchrotron spectral 'hardening', $\Delta \alpha$, as a function
of $b$. Model parameters: $b_{max}=\infty$, and $q=10, 10^2, 10^3$.}
\label{fig8}
\end{figure}

\clearpage

\begin{figure}
\centering
\includegraphics[width=5in,angle=0]{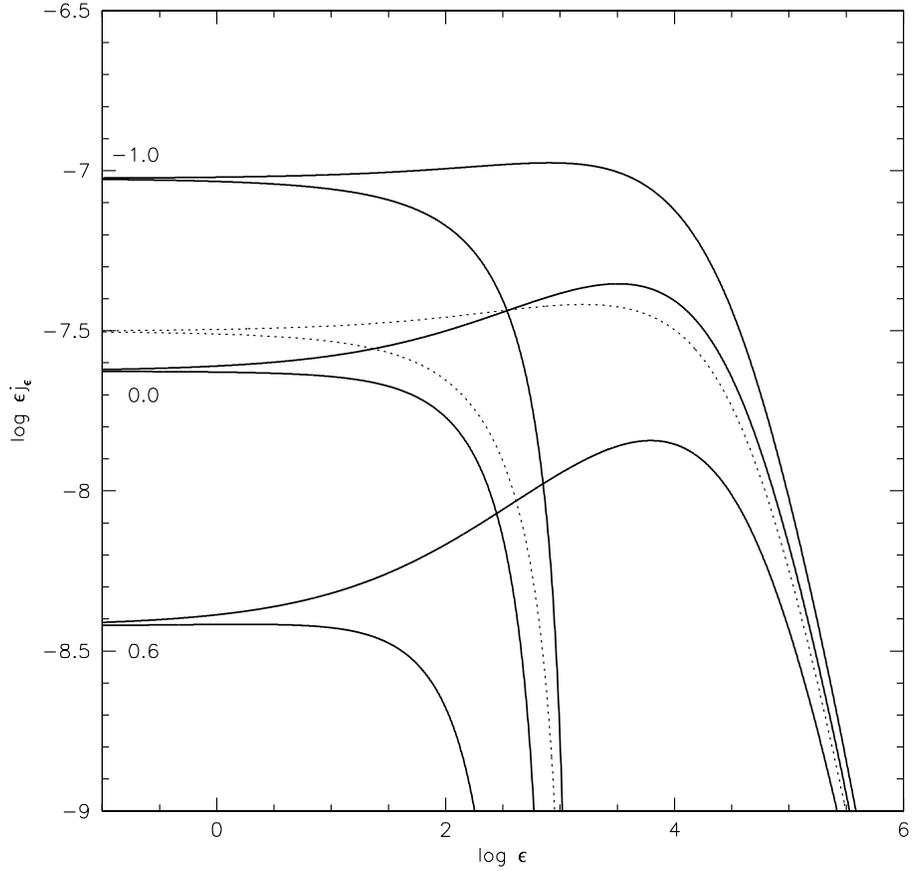}
\caption{Inverse-Compton spectra for a beamed ambient radiation field.
The model parameters: $q=30$, $b_{max}=10, 10^3$; $\epsilon_0=10^{-4}$;
$\cos \theta= -1.0, 0.0, 0.6$. For comparison, the IC spectra for an isotropic
ambient radiation field are also shown (dotted lines).}
\label{fig9}
\end{figure}

\begin{figure}
\centering
\includegraphics[width=5in,angle=0]{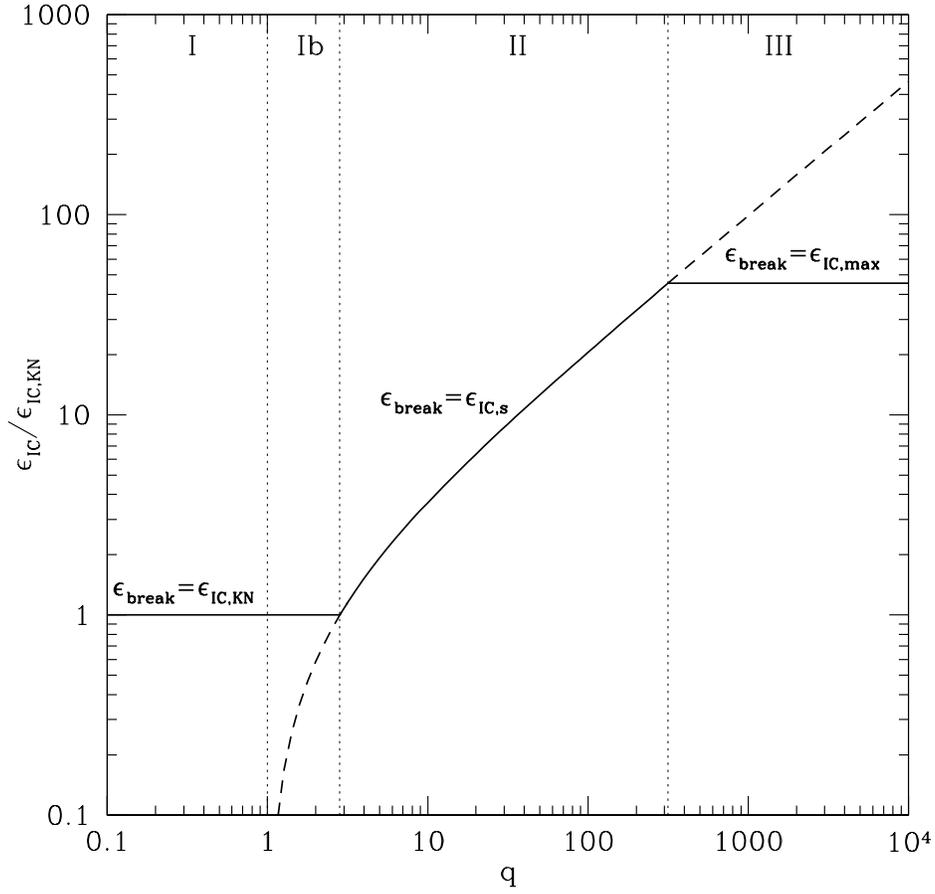}
\caption{
Schematic illustration showing the location of the high-energy 
break in the Compton spectra as a function of $q.$ (For details, see
text.) 
}
\label{fig10}
\end{figure}

\end{document}